\documentclass[
reprint,
amsmath,
amssymb,
aps]{revtex4-2}

\usepackage{graphicx}% Include figure files
\usepackage{bm}% bold math
\usepackage{hyperref}
\usepackage{braket}
\usepackage{makecell}

\newcommand{\him}[1]{{\ensuremath{\rm #1}}}

\begin{document}
	\preprint{APS/123-QED}
	\title{Statistical parameter estimation of multimode multiphoton subtracted thermal states of light}
	\author{G.~V.~Avosopiants}\email{avosopyantsgrant@gmail.com}
	\affiliation{Quantum Technology Centre, Faculty of Physics, M. V. Lomonosov Moscow State University,119991, Moscow, Russia}
	\author{B.~I.~Bantysh}
	\affiliation{Valiev Institute of Physics and Technology, Russian Academy of Sciences,117218, Moscow, Russia}
	\author{K.~G.~Katamadze}
	\affiliation{Quantum Technology Centre, Faculty of Physics, M. V. Lomonosov Moscow State University,119991, Moscow, Russia}
	\affiliation{Valiev Institute of Physics and Technology, Russian Academy of Sciences,117218, Moscow, Russia}
	\author{N.~A.~Bogdanova}
	\affiliation{Valiev Institute of Physics and Technology, Russian Academy of Sciences,117218, Moscow, Russia}
	\author{Yu.~I.~Bogdanov}
	\affiliation{Valiev Institute of Physics and Technology, Russian Academy of Sciences,117218, Moscow, Russia}
	\author{S.~P.~Kulik}
	\affiliation{Quantum Technology Centre, Faculty of Physics, M. V. Lomonosov Moscow State University,119991, Moscow, Russia}
	%\date{\today}
	
	\begin{abstract}
		Thermal states of light are widely used in quantum optics for various quantum phenomena testing. Particularly, they can be utilized for characterization of photon creation and photon annihilation operations. During the last decade the problem of photon subtraction from multimode quantum states become of much significance. Therefore, in this work we present a technique for statistical parameter estimation of multimode multiphoton subtracted thermal states of light, which can be used for multimode photon annihilation test.
	\end{abstract}
	\maketitle
	\section{Introduction}
	Photon creation and annihilation operators are base elements of quantum optics. Despite the fact that they are non-Hermitian and non-unitary, they can be directly (but probabilistically) implemented \cite{Ourjoumtsev2006,NeergaardNielsen2006,Parigi2007,Zavatta2004}. Thereby we get a perfect toolbox allowing tests of basic commutation rules \cite{Parigi2007}, Schrödinger's cat and other non-gaussian quantum state preparation \cite{Ourjoumtsev2006,NeergaardNielsen2006,Wenger2004}, probabilistic linear noiseless amplification \cite{Xiang2010}, strong Kerr nonlinearity implementation \cite{Costanzo2017}, etc.
	
	Thermal states of light are easy to prepare and its statistics is modified significantly by both photon creation and annihilation. Therefore, photon-subtracted thermal states (PSTS) become very attractive for the demonstration of effects in quantum optics and quantum thermodynamics like quantum vampire effect \cite{Katamadze2019a,Bogdanov2018}, photonic Maxwell's demon \cite{Vidrighin2016}, quantum thermal engine \cite{Hlousek2017}, etc. Moreover, it was shown, that PSTSs can be utilized in some metrological applications \cite{Parazzoli2016,Boyd2017}.
	
	Recently, the action of non-gaussian operations (particularly photon creation and annihilation) on the multimode states of light has become very interesting in the context of cluster-state quantum computing \cite{Andersen2015,Ra2020}. Despite the fact that there are some mode-selective photon subtraction techniques \cite{Ra2020,Ra2017}, generally the annihilation operator is implemented using a low-reflective beam splitter and a photon detector in the reflected channel \cite{Ourjoumtsev2006,NeergaardNielsen2006,Parigi2007}. In this case we cannot control in which optical mode the photon is subtracted. 
	
\begin{figure}[t]
		\center{\includegraphics[width=0.8\columnwidth]{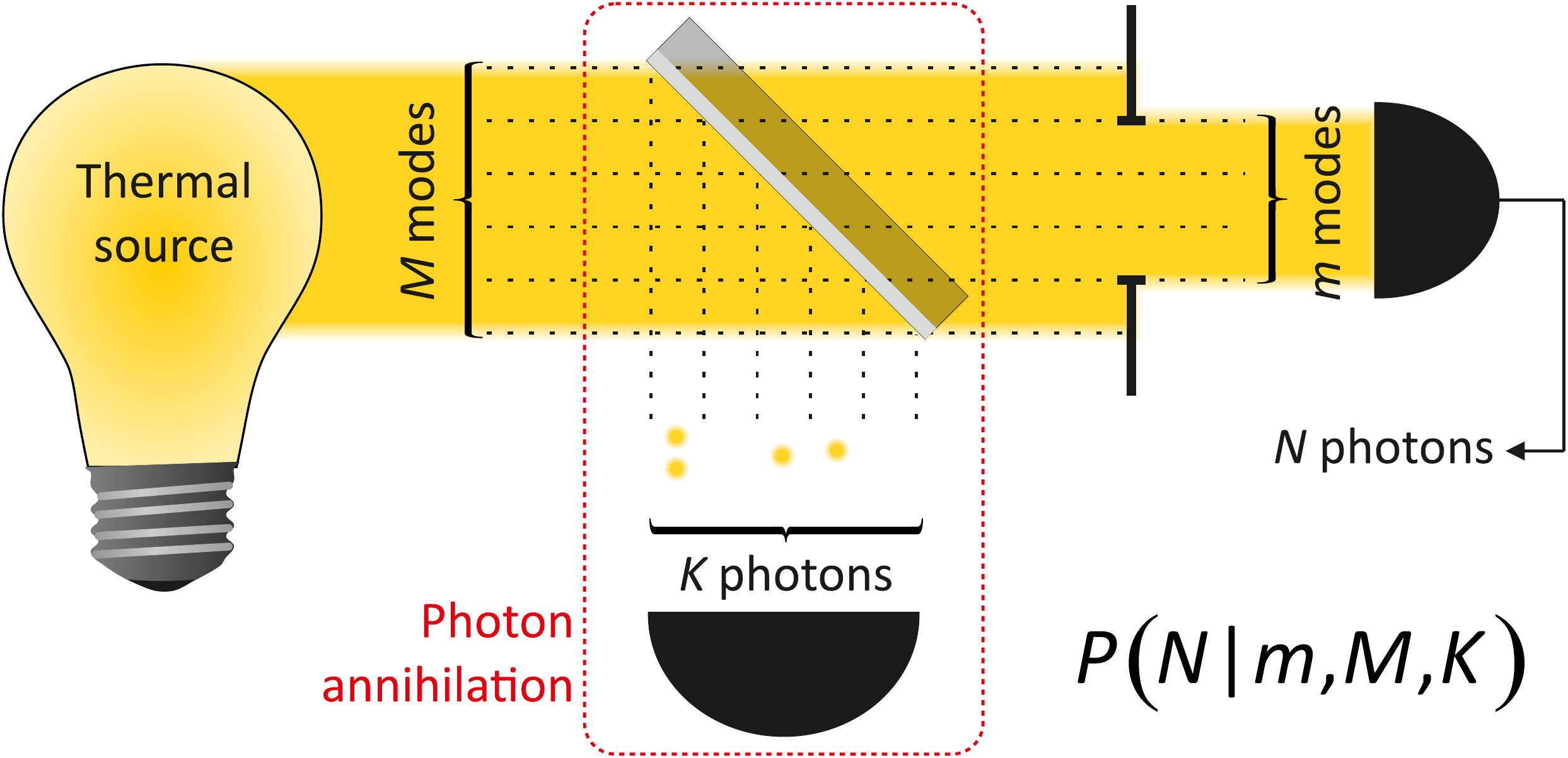}}
		\caption{(Color online). The registration scheme for photon number statistics of multiphoton-subtracted multimode thermal state.}
		\label{fig:1}
	\end{figure}

	In this work, we consider the photon subtraction from a multimode quantum state, and study only a part of the output modes (Fig.~\ref{fig:1}). The case of a single mode detection is of particular interest, e.g. the homodyne detection. We study an example of a multimode thermal state at the input and the general case of multiple photon detection in the reflected channel that corresponds to multiple photon subtraction.  

	Previously, it has been theoretically shown \cite{Agarwal1992} and experimentally verified \cite{Allevi2010,Zhai2013,Bogdanov2017} that the photon number distribution of $ K $-photon subtracted $ M $-mode thermal state can be described by a negative binomial, or a compound Poisson distribution \cite{Bogdanov2003} $ P_{cP}(N|\mu_0,a) $ with two parameters: the group parameter $ a=K+M $ and the initial per mode mean photon number $ \mu_0 $ \cite{Agarwal1992,Bogdanov2017,Bogdanov2016a,Katamadze2019,Mandel1995}:
	\begin{equation}\label{eq:Pcp}
		P_{cP}(N|\mu_0,a)=\frac{\Gamma(a+N)}{\Gamma(a)}\frac{\mu^N_0}{N!}\left(\frac{1}{1+\mu_0} \right)^{N+a}. 
	\end{equation}
	The mean photon number of this distribution is ${\mu=\mu_0 a}$.
	
	Thus, it is impossible to determine separately the number of modes $ M $ and the number of subtracted photons $ K $ by examining the total photocount statistics in all modes. One can only get the sum of these parameters. This creates problems in the case when we need to determine the parameters of multimode PSTS.
	
	But the situation changes somewhat if we consider only ${m<M}$ modes of the state described by the distribution (\ref{eq:Pcp}). Then the resulting photocount distribution in such a subsystem is the convolution of the compound Poisson distribution (\ref{eq:Pcp}) and the Polya distribution $ P_{Polya}(k|m,M,K) $:
	\begin{eqnarray}\label{eq:Pnkm}
		P(N|\mu_0,m,M,K)=\sum_{k=0}^{K}&&P_{Polya}(k|m,M,K)\times \nonumber\\
		&&\times P_{cP}(N|\mu_0,a=k+M),
	\end{eqnarray}
	where $ P_{Polya}(k|m,M,K)=\frac{C^k_{m+k-1}C^{K-k}_{M-m+K-k-1}}{C^K_{M+K-1}} $ \cite{Bogdanov2003,Avosopiants2019,Katamadze2020,Landau1959}.
	
	The convolution in (\ref{eq:Pnkm}) is easy to calculate using the generating functions approach. The generating function of the distribution (\ref{eq:Pnkm}) is following \cite{Feller1968}:
	\begin{eqnarray}\label{eq:Gzkm}
		G(z|\mu_0,m,M,&& K)=[G_{BE}(z|\mu_0)]^m \times \nonumber\\
		&&\times {}_2F_1\left(-K,m,M,1-G_{BE}(z|\mu_0)\right).
	\end{eqnarray}
	Here $ G_{BE}(z|\mu_0)=[1+\mu_0(1-z)]^{-1} $ is the generating function of the thermal state (Bose-Einstein distribution), $ {}_2F_1 $ is the Gaussian hypergeometric function (for more details see our work \cite{Katamadze2020}). The corresponding photon number statistics is
	\begin{eqnarray}\label{eq:Pfull}
		P&&(N|\mu_0,m,M,K)=\frac{\mu_0^N}{(1+\mu_0)^{N+m}}\times \nonumber\\
		&&\times\frac{1}{\Gamma(m)}\frac{\Gamma(N+m)}{\Gamma(N+1)}\frac{\Gamma(M)}{\Gamma(M-m)}\frac{\Gamma(M+K-m)}{\Gamma(M+K)}\times \nonumber\\
		&&\times{}_2F_1\left(-K,N+m,-K-M+m+1,\frac{1}{1+\mu_0}\right).
	\end{eqnarray} 
	
	In the previous work \cite{Katamadze2020}, we have shown that model (\ref{eq:Pfull}) is adequate to the experimental data, provided that all parameters are fixed except for the per mode mean photon number $ \mu_0 $ that was calculated from the experimental data. In this work, we investigate the possibility of estimating parameters $m$, $M$, $K$ and $\mu_0$ using model (\ref{eq:Pfull}) and the photocount statistics or the results of quadrature measurements. In the latter case, since the homodyne selects exactly a single mode, corresponding to the local oscillator, $ m=1 $.
	
	The paper has the following structure. Section \ref{sect:experiment} describes the procedures for preparation and measurement of various states of light of the form (\ref{eq:Pfull}). Section \ref{sect:est_photocount} describes the procedure for the statistical estimation of the state parameters based on measurements of the photocount statistics. The statistical estimation in Section \ref{sect:est_quadrature} is based on the quadrature measurements. We conclude that it is possible to use the model (\ref{eq:Pfull}) for statistical estimation of the parameters of multimode PSTS, if a prior information is provided.
	
	\section{Experiment}\label{sect:experiment}
\begin{figure}[t]
		\center{\includegraphics[width=0.8\columnwidth]{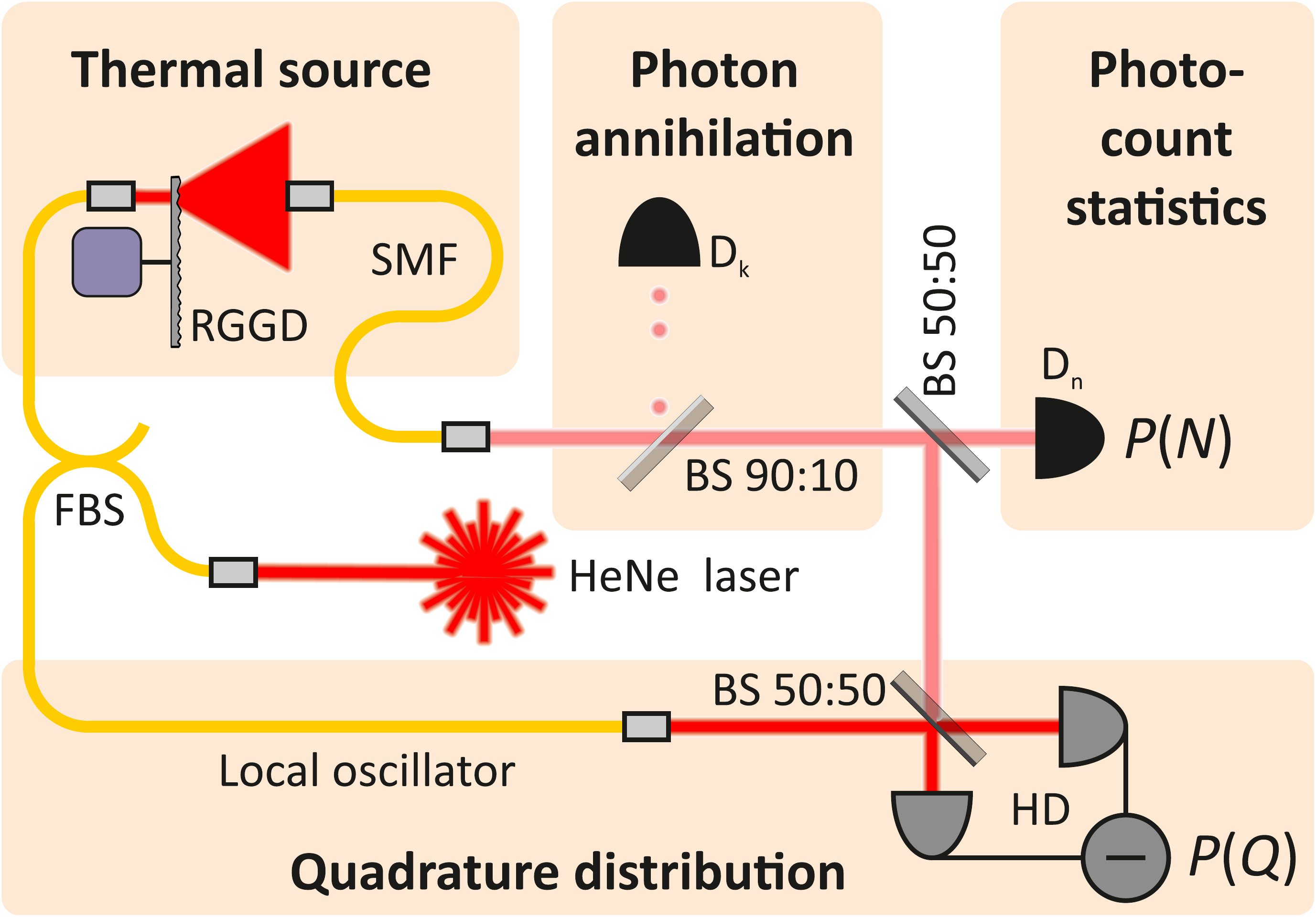}}
		\caption{(Color online). The experimental setup. BS --- beam splitters, FBS --- fiber-based beam splitter, RGGD --- rotating ground glass disk, SMF --- single-mode fiber, D\textsubscript{k} and D\textsubscript{n} are single-photon APD-based detectors used for photon annihilation and photocount statistics measurements respectively, HD --- homodyne detector, used for quadrature distribution registration.}
		\label{fig:2}
	\end{figure}

	The sketch of our experimental setup is presented in Fig.~\ref{fig:2}. The optical scheme represents the combination of schemes described in \cite{Bogdanov2017,Katamadze2020}. The HeNe cw laser beam is split by a fiber beam splitter (FBS) into two channels. The light from the first output is focused on a rotated ground glass disk (RGGD) and a part of the scattered light is coupled into a single-mode fiber (SMF) for the single-mode thermal state preparation \cite{Martienssen1964,Arecchi1965}. A small part of the fiber output beam is redirected by a 90:10 beam splitter (BS) to a single-photon detector D\textsubscript{k} based on a silicon APD, in order to implement conditional photon annihilation \cite{Ourjoumtsev2006,NeergaardNielsen2006}. Next the radiation is split by a symmetric BS into two parts. In the first one there is an another APD detector D\textsubscript{n} for photocount distribution measurement. The rest of the beam is subjected to the homodyne detection HD. Laser beam from the second output of the FBS serves as a homodyne local oscillator. Since the quadrature distribution of thermal states, as well as MPSTSs, does not depend on the homodyne phase, the phase didn't fixed. Thus, photocount pulses from D\textsubscript{n} and D\textsubscript{k} and quadrature values from HD are collected synchronously. It allows to study the photon statistics registered by the detector D\textsubscript{n} and the quadrature statistics obtained by the HD under the condition of a given number of subtracted photons, collected by the detector D\textsubscript{k}.
	
	It is important to note that splitting the radiation in half inevitably leads to losses. However, the PSTSs described by the compound Poisson distribution $P_{cP}(N|\mu_0,a)$ under the influence of losses converts to the PSTSs $P_{cP}(N|\mu_0^\prime,a)$ with a lower mean photon number $\mu_0^\prime$, but with the same $a$ parameter \cite{Bogdanov2017,Bogdanov2016a}.
	
\begin{figure}[t]
		\center{\includegraphics[width=0.8\columnwidth]{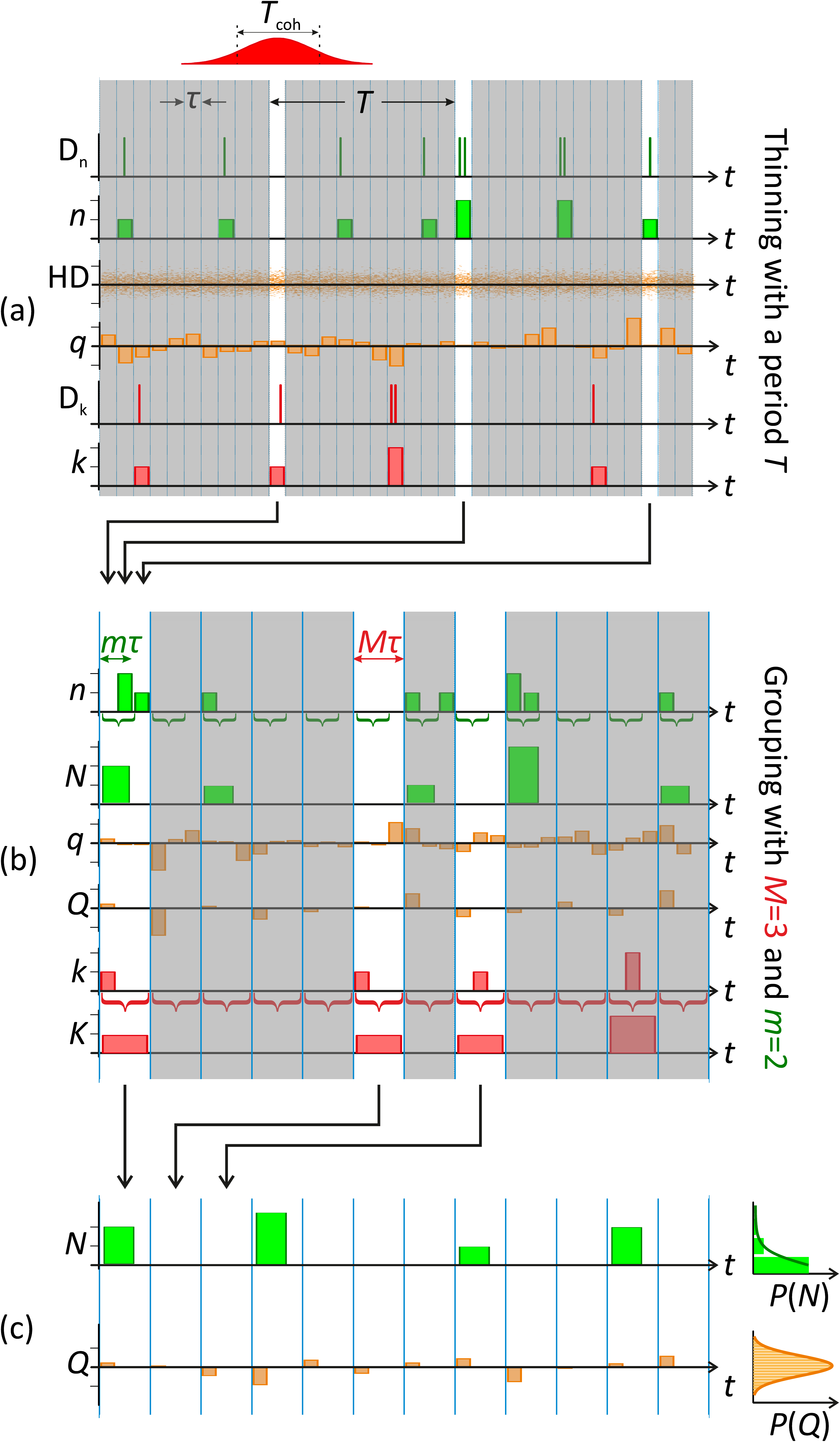}}
		\caption{(Color online). Signal processing. (a) Initial data set is divided into time bins $ \tau $ and then they are thinned with a period $ T $ in order to avoid interbin correlations. (b) Thinned data are grouped by $ M $ and groups are separated according to the total number of subtracted photons in the group $ K $. The total photon number $ N $ is calculated as a sum of the first $ m\leq M $ bins. For a group quadrature value $ Q $ the first bin value $ q $ is selected. (c) The data sets ${\{N_1,N_2,\dots\}}$ and ${\{Q_1,Q_2,\dots\}}$ corresponded to the same value of $ K $ are collected and subjected to the statistical estimation procedures.}
		\label{fig:3}
	\end{figure}

	The data processing algorithm is presented in Fig.~\ref{fig:3}. First, all the time traces are divided into time bins with the width $ \tau $  corresponding to the time mode duration (Fig.~\ref{fig:3}a). The value of $ \tau $ should satisfy the inequality ${T_{coh}\gg \tau \gg \tau_d}$, where $ T_{coh} $ is the thermal state coherence time defined by the RGGD velocity and $ \tau_d $ is the single-photon detector dead time. This inequality defines the possibility of several photocounts registration from a single optical mode (see \cite{Bogdanov2017} for details). In our experiment $ T_{coh}=40~\him{\mu s} $, $ \tau_d=220~\him{ns} $ and $ \tau=10~\him{\mu s} $, so the inequality is satisfied and we were able to register up to 45 photons in each time bin. For each bin the photocount numbers $ k $ and $ n $ from the detectors D\textsubscript{k} and D\textsubscript{n} respectively and the quadrature values $q$ from HD are calculated. Next, in order to avoid any interbin correlations, we selected the bins periodically separated by $ T=12T_{coh} $. Such a large interval $T$ is necessary, since the thermal field has a Gaussian correlation function, and even at times significantly longer than the coherence time $T_{coh}$, correlations are partially preserved, which distorts the photon number and quadrature statistics. Thus, only 2\% of the collected data is used.
	
	In contrast to the situation considered in Fig.~\ref{fig:1}, where various spatial modes of the thermal field were considered, in our experiment the field is spatially single-mode, so one can select a multimode state by collecting $ M $ time modes. Therefore, all the uncorrelated time bins are grouped by $ M $ (Fig.~\ref{fig:3}b). For each group we obtain the total number of subtracted photons $ K $. In order to realize the situation described in Fig.~\ref{fig:1}, where just a part of the thermal modes is finally collected, we calculate the total photon number $ N $ as a sum of the first $ m $ bins in a group. Since homodyne can select only a single mode, we take only the first bin quadrature value $  q $ in a group as a group quadrature value $ Q $. This value corresponds to the single mode quadrature value under the condition, that $K$ photons have been subtracted from the corresponding group of $M$ time modes. 
	
	To extract the $ K $-photon subtracted state we select the groups with the total number of annihilated photons equals $ K $ (Fig.\ref{fig:3}c). Thus, for each value of $ M=1 \div 5 $, $ m=1 \div M $ and $ K=0\div 5 $ we derive a set of photocounts values $ \mathcal{D}=\{N_1,N_2,\dots\} $ and quadrature values $ \mathcal{D}_Q=\{Q_1,Q_2,\dots\} $. These data sets are subsequently processed to reconstruct the state parameters using the distribution models (\ref{eq:Pfull}) and (\ref{eq:pq}) respectively.
		
	\begin{figure}[t]
		\center{\includegraphics[width=1\columnwidth]{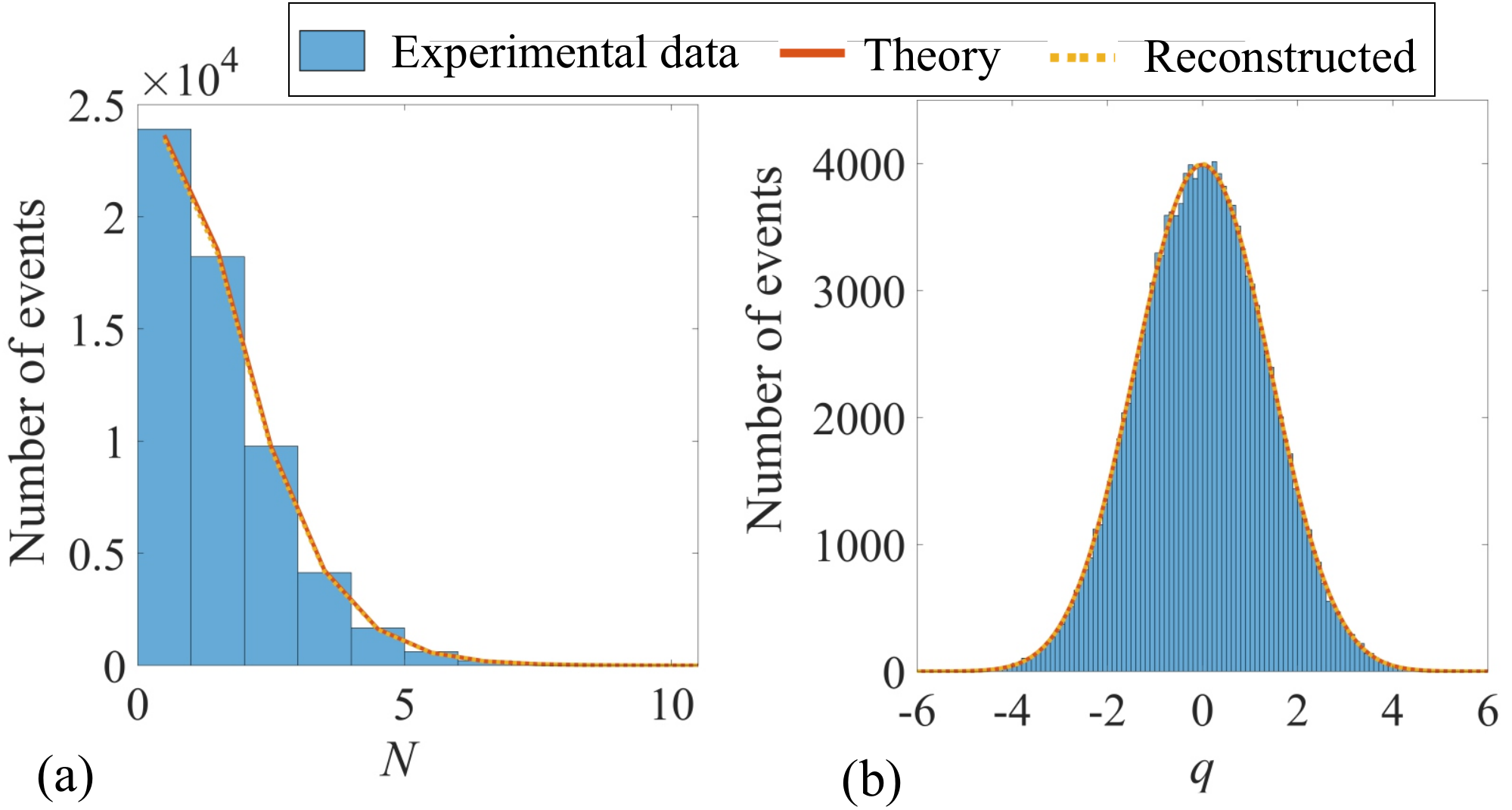}}
		\caption{(Color online). Experimental data (histograms) compared to probability distributions with theoretical (solid curves) and reconstructed (dashed curves) parameters values for photocount (a) and quadrature (b) statistics.}
		\label{fig:4}
	\end{figure} 	
	
	Thus, we are able to extract the data for an arbitrary state of light with photocount distribution (\ref{eq:Pfull}) in a wide range of well controlled parameters $m$, $M$ and $K$. However, the setup does not allow us to control the per mode mean photon number $\mu_0$, so we estimate its theoretical value from the data using equation \cite{Katamadze2020}:
	\begin{equation}\label{eq:mu}
		\mu_0=\frac{\mu}{m\left(1+\frac{K}{M} \right) },
	\end{equation}
	where $ \mu $ is the estimated mean photon number in all registered modes.
	
	Note that $ P(N) $ does not exactly correspond to the experimental photocount distribution because of the presence of the dark counts, described by the Poisson distribution $ P_{DC}(N) $ with the mean value $ \mu_{DC}=m\times 0.0015 $. Despite the fact that the average number of noise photocounts is much less than the average number of photons per mode (about $ \mu_0=0.27 $ in examples below), we take it into account to increase the reconstruction accuracy. The resulting photocount distribution is the convolution of (\ref{eq:Pfull}) and $ P_{DC}(N) $.
	
	\section{Parameters estimation based on photocount statistics}\label{sect:est_photocount}
	In this section we examine in detail an example based on the experiment with theoretical parameters values ${m_t=2}$, ${M_t=3}$, ${K_t=3}$. The total number of observed events was $ n=58623 $. The calculated theoretical value of per mode mean photon number was ${\mu_{0,t}=0.264}$. We denote the number of $N$-photocounts events in the sample $\mathcal{D}$ as $D(N)$. The corresponding histogram and the probability distribution based on theoretical values are shown in Fig.~\ref{fig:4}a.
	
	First, we develop a parameters estimation procedure using simulated data. Then we apply it to process the real experimental data.
	
	\subsection{Multicollinearity}
\begin{figure}[t]
		\center{\includegraphics[width=1\columnwidth]{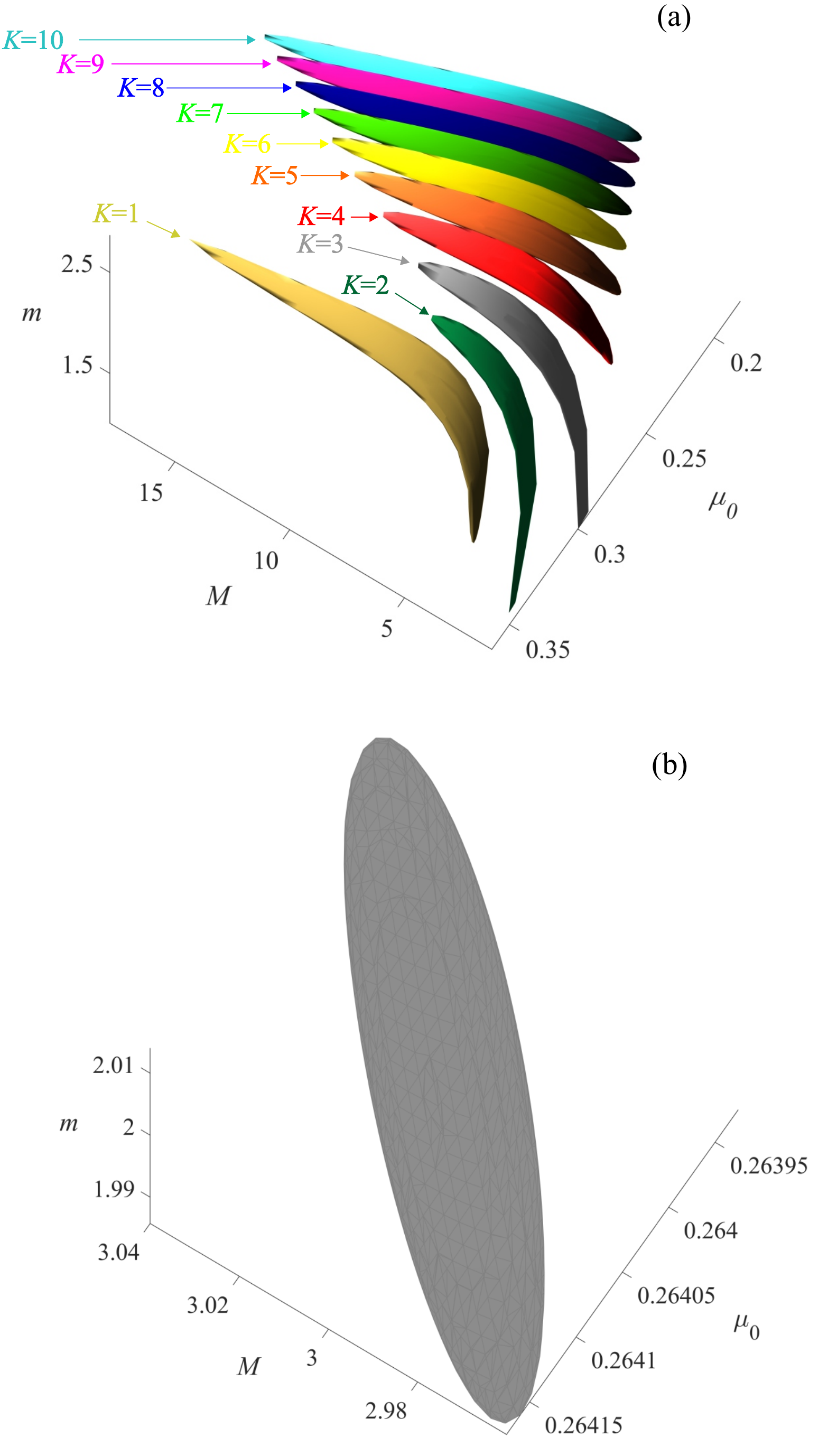}}
		\caption{(Color online). Isosurfaces of the fiducial distribution ${P_F(\mu_0,m,M,K|\mathcal{D})}$ at the half-maximum level for the fixed values of $K$. The data $\mathcal{D}$ was obtained using Monte Carlo simulation with sample size $n$ and distribution parameters ${\mu_{0,t}=0.264}$, ${m_t=2}$, ${M_t=3}$, ${K_t=3}$. (a) ${K=1\div 10}$, ${n = 58623}$. (b) ${K = 3}$, ${n=420\cdot 10^6}$.}
		\label{fig:5}
	\end{figure}

	Consider the fiducial distribution of parameters $ P_F(\mu_0,m,M,K|\mathcal{D}) $. One can interpret this distribution as the degree of confidence that a certain set of parameters $ \{\mu_0,m,M,K\} $ conditions the data set $ \mathcal{D} $. The distribution is equal to the likelihood function $ L $ up to the normalization constant $ C $ \cite{Fisher1935,Cox2006,Kendall1961}:
	\begin{equation}\label{eq:Lmkn}
		L(\mu_0,m,M,K|\mathcal{D})=\prod_{N=0,1,\ldots}\left[P(N|\mu_0,m,M,K) \right]^{D(N)}. 
	\end{equation}
	For the sample size ${n = 58623}$, the width of the marginal fiducial distributions over any parameter is quite large. The width of the marginal distribution over $ K $ is especially large estimating hundreds of units. This is due to strong correlations between the parameters, or multicollinearity of the initial distribution (\ref{eq:Pfull}). Fig.~\ref{fig:5}a well illustrates this effect.
	
	Note that the likelihood function takes very small values for a high sample size, so we consider its logarithm:
	\begin{eqnarray}\label{eq:lnLmkn}
		\ln L(\mu_0,&& m,M,K|\mathcal{D})= \nonumber \\
		&& =\sum_{N=0,1,\ldots}D(N)\ln P(N|\mu_0,m,M,K).
	\end{eqnarray}
	Since adding a constant to a given function only affects the proportionality constant $C$, it is efficient to calculate the fiducial distribution relative to the shifted logarithmic likelihood:
	\begin{eqnarray}\label{eq:pf}
		&& P_F(\mu_0,m,M,K|\mathcal{D})=C^{\prime} \cdot \nonumber\\
		&&\cdot \exp{\left[{\scriptstyle  \ln L(\mu_0,m,M,K|\mathcal{D})-\max\limits_{\mu_0,m,M,K}\ln L(\mu_0,m,M,K|\mathcal{D})} \right]}. 
	\end{eqnarray}
	As a result of this shift, the exponent values range from 0 to 1. The constant $ C^{\prime} $ is then calculated by the direct integration.
	
	To numerically characterize multicollinearity, one can calculate the Fisher information matrix ${I_{u,v}=n\mathbb{E}_N[(\partial_u P)(\partial_v P)]}$, where $ \partial_u P $ is the partial derivative of the distribution with respect to the parameter $ u $, and $ u,v=m,M,\mu_0 $. Here we assume the parameter $K$ to be fixed. According to the Cram\'er-Rao bound, the covariance matrix for the estimates of the distribution parameters is bounded by the reciprocal of the Fisher information $I^{-1}$ \cite{Kendall1961}. Thus, the condition number of the Fisher information matrix (the ratio between its maximum and minimum eigenvalues) reflects the robustness of statistical estimates with respect to statistical fluctuations. For all practically important parameters values considered in our study, the information matrix turns out to be ill-conditioned. In particular, for the above case the condition number is about 7 million. This results in a very low accuracy of statistical estimates.
	
	Note that the inverse Fisher information matrix gives the estimates variances only for a fixed value of $K$. Therefore, we characterize the parameters estimation accuracy by the maximum relative error ${\Delta=\max_{u}(\sigma_{u}/{u_t})}$ (${u=m,M,\mu_0,K}$) to take fluctuations of $K$ into account. Here $ \sigma_{u} $ is the marginal standard deviation of the fiducial distribution $ P_F(\mu_0,m,M,K|\mathcal{D}) $, and $u_t$ is the parameter theoretical value.
	
	The multicollinearity significantly complicates the procedure for the state parameters reconstruction. To obtain a sufficient reconstruction accuracy, a very large amount of data is required (which is difficult to implement for high $ K $ values, since they correspond to relatively rare events). For example, numerical experiments show that one needs a sample size of at least $ n=420\cdot 10^6 $ in order to achieve $ \Delta=1\% $ precision (Fig.~\ref{fig:5}b). To achieve $ \Delta=10\% $ one needs at least $ n=18\cdot 10^6 $ being still a large amount of data.
	
	\subsection{Prior information}
	Introducing some prior information could, in principle, increase the estimation accuracy.
	
\begin{figure*}[t]
		\center{\includegraphics[width=2\columnwidth]{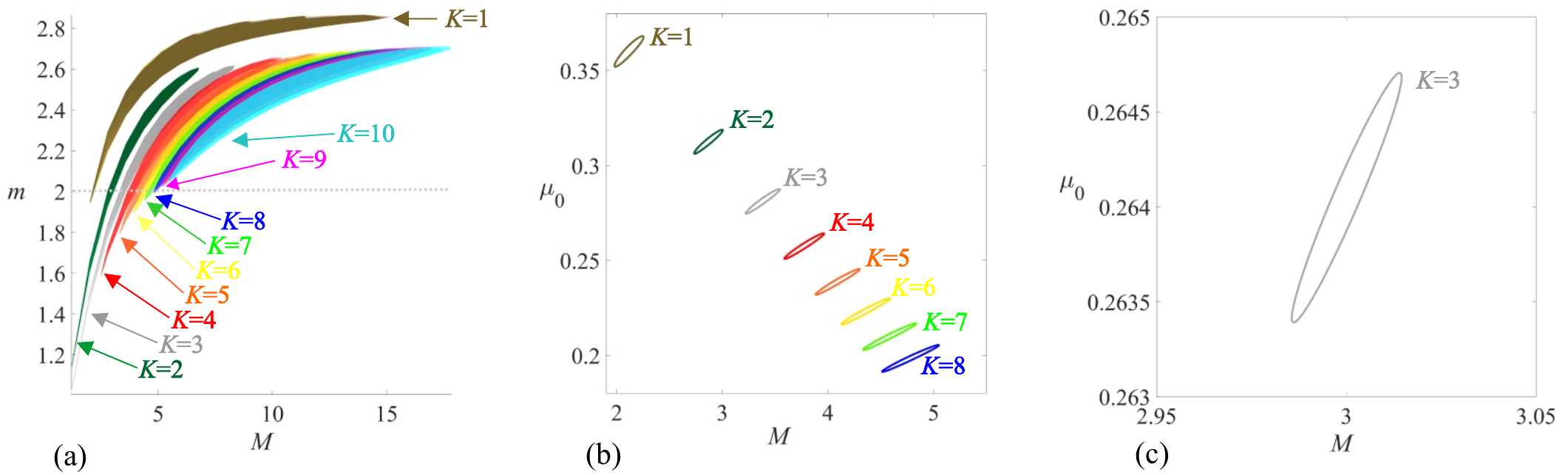}}
		\caption{(Color online). a) Projection of the fiducial distribution isosurfaces (Fig.~\ref{fig:5}a) onto the plane $ \{M,m\} $. The horizontal line corresponds to the plane with $ m = m_t $. b-c) Cross-sections of the fiducial distribution isosurfaces at $ m = m_t $ for $ n=58623 $ (b) and $ n=4\cdot 10^6 $ (c). The latter case provides a single plausible value of $ K $ $ (K=3) $ and $ \Delta=1\% $ relative error.}
		\label{fig:6}
	\end{figure*}

	A common choice is to fix the value of some parameter (or a set of parameters). In particular, one can control the number of selected modes $m$. For example, with homodyne detection, only a single mode of light is selected ($m_t = 1$). In this case, the fiducial distribution takes the form ${P^m_F(\mu_0,M,K|\mathcal{D})=C_m L(\mu_0,m=m_t,M,K|\mathcal{D})}$. Graphically, this corresponds to the plot cross-section at $ m = m_t $ (horizontal line in Fig.~\ref{fig:6}a) that intersects the distribution isosurfaces for $ K = 1 \div 8 $ only. Thus, fixing $ m $ reduces the number of plausible values of $ K $ from hundreds to the order of ten. However, corresponding cross-sections also show strong parameters correlations (Fig.~\ref{fig:6}b) and multicollinearity.
	
	Note that to achieve ${\Delta=10\%}$ one needs a sample size of at least ${n=1.2 \cdot 10^6}$. For ${\Delta=1\%}$ (Fig.~\ref{fig:6}c) the sample size ${n=4\cdot 10^6}$ is required. This is still a quite large amount of data, since the cases of high numbers of subtracted photons are less frequent and require more time to gather experimental data.
	
	\subsection{Bayesian inference}
	Fixing certain parameters of the initial distribution can significantly improve the reconstruction accuracy. This, however, can introduce systematic errors of reconstruction, if the selected prior values differ significantly from the true values.
	
	Another approach of using prior information is based on the Bayes' theorem \cite{Feller1968,Gelman2013}:
	\begin{eqnarray}\label{eq:pb}
		P_B(\mu_0,m,M,K|\mathcal{D})=C_B L(\mu_0&&,m,M,K|\mathcal{D})\cdot \nonumber \\
		&& \cdot P_P(\mu_0,m,M,K).
	\end{eqnarray}
	Here $ P_P(\mu_0,m,M,K) $ is the prior probability distribution of plausible parameters values. The posterior distribution $ P_B(\mu_0,m,M,K|\mathcal{D}) $ updates the prior information, taking into account the statistical data $ \mathcal{D} $ obtained in the experiment.
	
	We rely on the common choice of a multi-parameter prior distribution, where all parameters are independent:
	\begin{eqnarray}\label{eq:pp}
		P_P(\mu_0,m,M,K)= P^{\mu_0}_P(\mu_0)&&\cdot P^m_P(m) \cdot \nonumber \\
			&& \cdot P^M_P(M) \cdot P^K_P(K).
	\end{eqnarray}
	To get single-parameter prior distributions, we consider the conditional distributions: all parameters, except one, are fixed and equal to the expected theoretical values. Further, we will demonstrate that such conditional distributions adequately describe our experimental data.
	
	\subsection{Conditional distribution verification}
\begin{figure}[t]
		\center{\includegraphics[width=1\columnwidth]{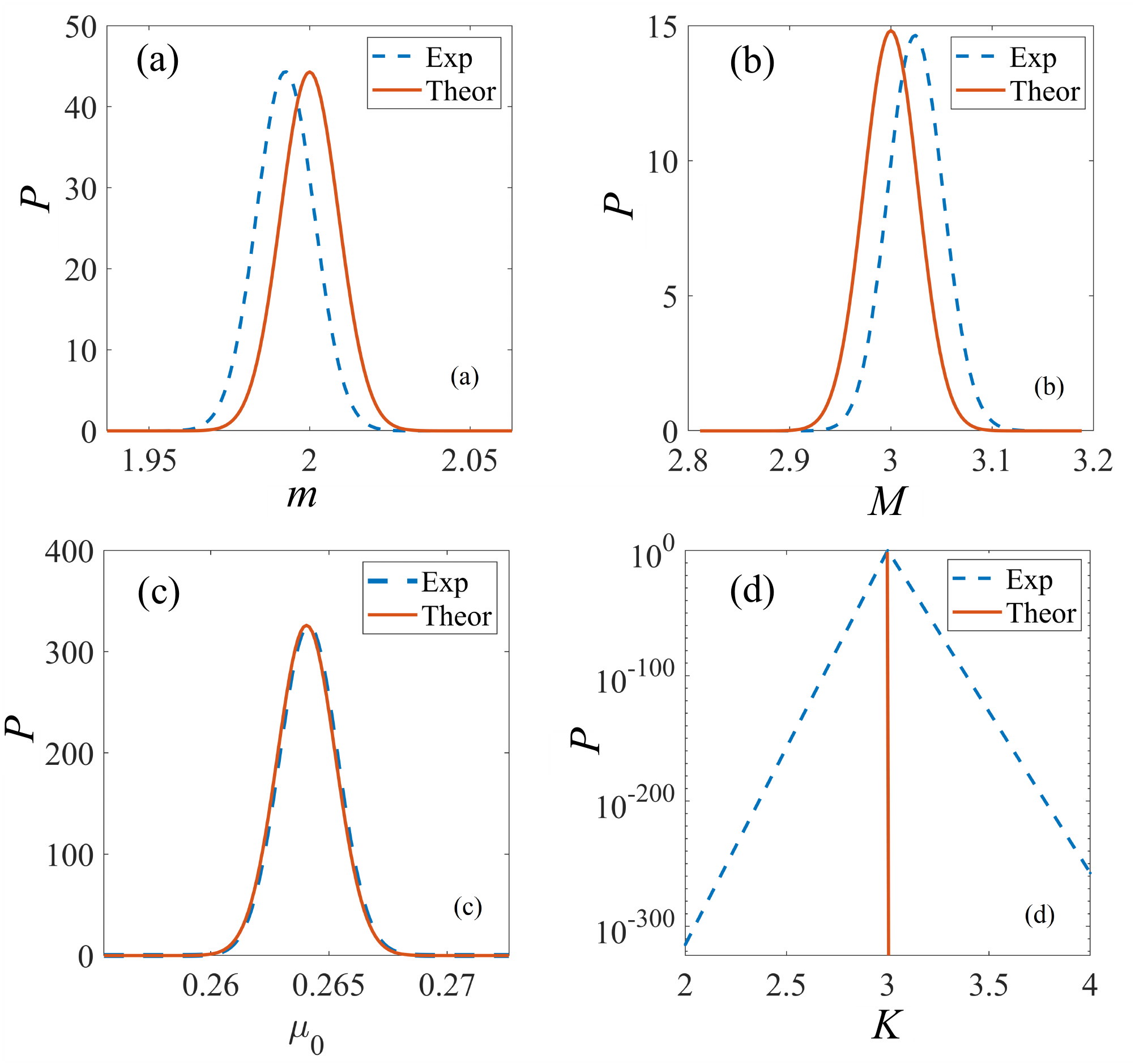}}
		\caption{(Color online). Sample (experimental) and exact (theoretical) Conditional fiducial distributions of the parameters $ m $ (a), $ M $ (b), $ \mu_0 $ (c) and $ K $ (d) based on experimental (dashed lines) and theoretical (solid lines) data.}
		\label{fig:7}
	\end{figure}

	Consider the conditional distribution for the parameter $ m $. Let us construct (Fig.~\ref{fig:7}a) two fiducial distributions: ${P_F^m(m|\mathcal{D})}$ and ${P_F^m(m|\mathcal{D}_t)}$, where ${P_F^m(m|\mathcal{D})=P_F(\mu_{0,t},m,M_t,K_t)}$ and ``data'' $\mathcal{D}_t$ corresponds to the theoretical grouped data ${D_t(N) = nP(N|\mu_{0,t},m_t,M_t,K_t)}$. Fig.~\ref{fig:7}a shows a strong overlap between these distributions, which suggests that the conditional distribution can be used to describe the data $ \mathcal{D} $. A similar result for parameters $ M $, $ \mu_0 $ and $ K $ are shown in Fig.~\ref{fig:7}b,c,d respectively. Note that, as follows from Fig.~\ref{fig:7}d, the parameter $ K $ is in fact deterministic, since the fiducial probability of $ K\neq K_t $ is almost zero. In this regard, below we consider $ P^K_P(K)=\delta_{K,K_t} $.
	
	\subsection{Prior distributions}\label{sect:prior_distr}
	Considering separately the conditional distributions introduced above, it is possible to perform the parameter reconstruction by the maximum likelihood estimation (MLE) technique. Let us take the parameter $ m $ as an example. We denote its MLE value as $\hat{m}_c$. Here and below the subscript $c$ stands for the estimates based on conditional distributions.
	
	According to the general estimation theory, in the limit of a high sample size $n$, MLE value is the random variable with normal distribution $f(\hat{m}|m_c,\sigma_{m,c})$ \cite{Kendall1961}. The expected value $m_c$ corresponds to the asymptotic ($n\rightarrow\infty$) MLE estimate and the variance is related to the single-parameter Fisher information: ${\sigma^2_{m,c}=I^{-1}_{mm}}$. Again, we refer to the fiducial inference denoting ${P^m_P(m)=f(m|\hat{m}_c,\sigma_{m,c})}$ as the prior distribution of the parameter $m$. The prior distributions $ P^M_P(M) $ and $ P^{\mu_0}_P(\mu_0) $ are calculated in a similar way.
	
	Following the above technique, we estimated the prior distribution paramaters for the real experimental data $\mathcal{D}$: $ \hat{m}_c=1.993 $, $ \sigma_{m,c}=0.009 $, $ \hat{M}_c=3.026 $, $ \sigma_{m,c}=0.027 $, $ \hat{\mu}_{0,c}=0.265 $, $ \sigma_{\mu_0,c}=0.001 $. One can observe a close relation between the MLE values and theoretical values. This again gives an evidence that the theoretical values used to build conditional distributions do not introduce an observable estimator bias.
	
	\subsection{Posterior distribution}
\begin{figure}[t]
		\center{\includegraphics[width=0.8\columnwidth]{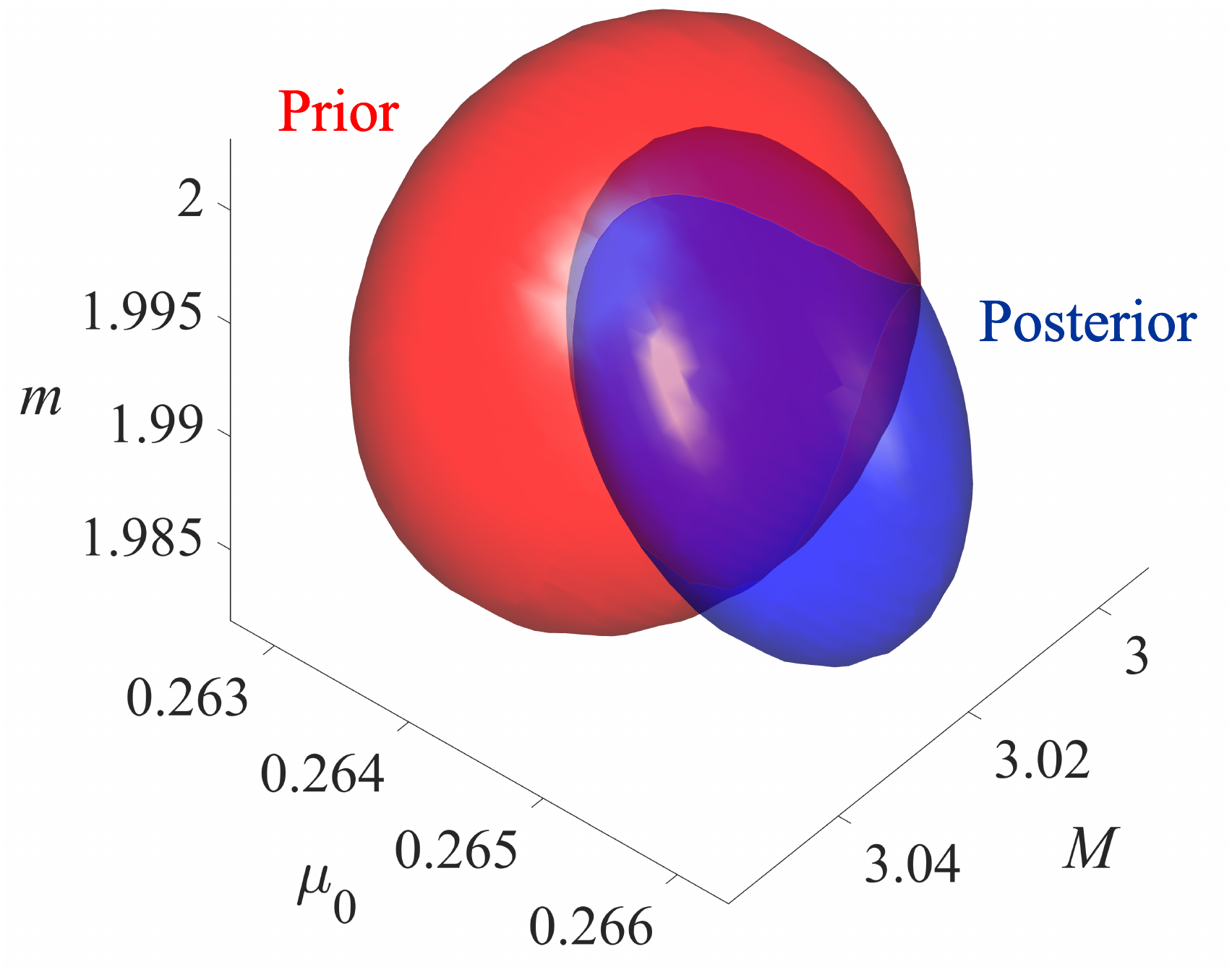}}
		\caption{(Color online). Half-maximum level isosurfaces of the prior $ P_P(\mu_0,m,M,K) $ and posterior $ P_F(\mu_0,m,M,K|\mathcal{D}) $ distributions of plausible parameters values.}
		\label{fig:8}
	\end{figure}

	Above, we obtained single-parameter prior distributions of all parameters under consideration. Their product forms the multi-parameter prior distribution (\ref{eq:pp}). The posterior distribution is the product of the multi-parameter distribution (\ref{eq:pp}) and the unconditional fiducial distribution (normalized likelihood function). Prior and posterior distribution are illustrated in Fig.~\ref{fig:8}.
	
	We calculate the Bayesian posterior distribution expected values to get the point estimates. The obtained values were close to the theoretical ones: $ \hat{m}_B=1.943 $, $ \hat{M}_B=3.084 $, $ \hat{\mu}_{0,B}=0.274 $. Fig.~\ref{fig:4}a shows (dashed curve) the distribution (\ref{eq:Pfull}) with these parameters values. The resulting curve is in a close agreement with the curve corresponding to the theoretical parameters.
	
	Thus, our approach implies using pre-determined theoretical values of parameters as the staring point for the prior distribution definition. These values are then clarified by means of Bayes' theorem taking statistical data into account. 
	
	To show that this approach avoids the multicollinearity problem, we construct an information matrix corresponding to the posterior distribution. First of all, note that the Fisher information of the normal distribution $f(\hat{m}|m_c,\sigma_{m,c})$ is ${1/\sigma^2_{m,c}=I_{mm}}$ (similarly for the other two parameters). Since Fisher information matrix of the product of probability distributions is the sum of the corresponding information matrices for each distribution separately, we finally get
	\begin{equation}\label{eq:fisher}
		I_B=I+I_P,
	\end{equation}
	where
	\begin{equation}\label{eq:fishM}
		I_P=		
		\begin{pmatrix}
			\frac{1}{\sigma^2_{m,c}} & 0 & 0 \\
			0 & \frac{1}{\sigma^2_{M,c}} & 0 \\
			0 & 0 & \frac{1}{\sigma^2_{\mu_0,c}}
		\end{pmatrix}
		=
		\begin{pmatrix}
			I_{mm} & 0 & 0 \\
			0 & I_{MM} & 0 \\
			0 & 0 & I_{\mu_0 \mu_0}
		\end{pmatrix}
	.
	\end{equation}

	In fact, the use of the prior information in the form of the product of normal fiducial distributions of MLE estimates doubles the diagonal of the information matrix of the original unconditional distribution. In our case,  the condition number of the resulting matrix is 750. This value is significantly lower than the condition number 7 million for unconditional distribution. Hence, the resulting covariance matrix gives low parameters estimator variances. Even for the sample of a low size, which was available in our experiment, one could obtain error rate below $ \Delta=1\% $.
	
	\begin{table}[b]%The best place to locate the table environment is directly after its first reference in text
		\caption{The sample size values required to achieve error rates below 1\% and 10\% for different reconstruction methods. \label{tab:1}}
		\begin{ruledtabular}
			\begin{tabular}{cccc}
				\thead{} & \thead{No prior \\ information} & \thead{Fixed m} & \thead{Bayesian \\ inference} \\
				\colrule
				$ \Delta=10\% $ & $ 18\cdot 10^6 $ & $ 1.2\cdot 10^6 $ & $ 8\cdot 10^2 $\\
				$ \Delta=1\% $ & $ 42\cdot 10^7 $ & $ 4\cdot 10^6 $ & $ 5.8\cdot 10^4 $
			\end{tabular}
		\end{ruledtabular}
	\end{table}
	
	Table~\ref{tab:1} shows the numerical characteristics obtained for the simulated photocount statistics. We derive the sample size required to achieve the error rate $ \Delta=1\% $ and $ \Delta=10\% $. We compare methods differ in types of prior information: the absence of any prior information, a known fixed value of $ m $, and the approximate knowledge of theoretical parameters values. The table~\ref{tab:1} clearly shows enormous amounts of data required to defeat multicollinearity. While the Bayesian method allows one to get an error rate below $ \Delta=1\% $ with the amount of data accumulated in the real experiment.
	
	Using the Bayesian inference, we estimated parameters of 90 different states with $ M=1 \div 5 $, $ m=1\div M $ and $ K=0 \div 5 $. The estimation error rate was from 0.0008\% to 0.03\% for $ \mu_0 $, from 0.002\% to 1.09\% for $ m $, and from 0.015\% to 1.89\% for $ M $.
	
	\section{Parameters estimation based on quadrature measurements}\label{sect:est_quadrature}
	A homodyne detector selects only a single-mode subsystem of the state ($m=1$). Let us consider the measurements results of the state with $ M_t = 5 $, $ K_t = 4 $. The data $\mathcal{D}_Q$ size was $ n = 138710 $ (Fig.~\ref{fig:4}b). Using (\ref{eq:mu}) with $ \mu=\sigma^2_q-1/2 $, where $ \sigma^2_q $ is the sample quadrature variance, we estimate $ \mu_{0,t}=0.752 $.
	
	The transition from the photocount distribution to the quadrature distribution of the electromagnetic field is carried out as follows \cite{Leonhardt1997,Bogdanov2016a,Bogdanov2016}:
	\begin{equation}\label{eq:pq}
		\tilde{P}(Q|\mu_0,M,K)=\sum_{N}P(N|\mu_0,m=1,M,K)|\varphi_N(Q)|^2,
	\end{equation}
	where $ \varphi_N(Q) $ are the eigenfunctions of a harmonic oscillator. These functions have the form of Chebyshev--Hermite basis. The explicit form of these functions is following:
	\begin{equation}\label{eq:osc}
		\varphi_N(Q)=\frac{1}{(2^N N! \sqrt{\pi})^{\frac{1}{2}}} H_N(Q) \exp\left(-\frac{Q^2}{2} \right),
	\end{equation}
	where $H_N(Q)$ is the $N$-th Hermite polynomial.
	
	We use the same technique as in the previous section to estimate the parameters of (\ref{eq:Pfull}) from $\mathcal{D}_Q$ (Fig.~\ref{fig:4}b). To begin with, we construct a fiducial distribution ${{\tilde{P}_F(\mu_0,M,K|\mathcal{D}_Q)=C \cdot  L(\mu_0,M,K|\mathcal{D}_Q)}}$, where the likelihood function has the form ${L(\mu_0,M,K|\mathcal{D}_Q)=\prod_{i}\tilde{P}(Q_i|\mu_0,M,K)}$.
	
	\begin{figure}[t]
		\center{\includegraphics[width=0.8\columnwidth]{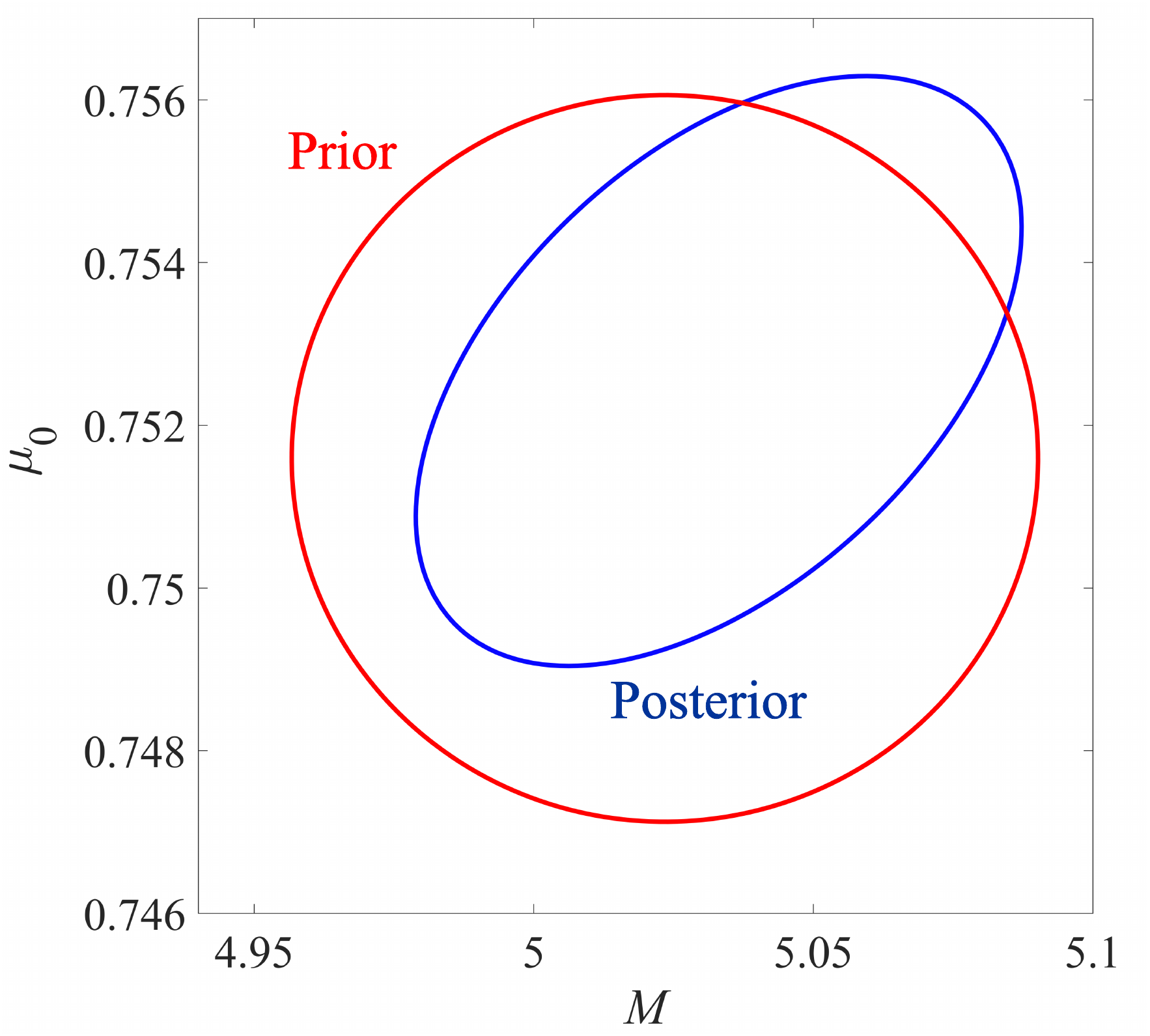}}
		\caption{(Color online). Half-maximum level isosurfaces of the prior and posterior distributions of plausible parameters values.}
		\label{fig:9}
	\end{figure}

	We again obtain a deterministic value ${K=K_t=4}$ and use theoretical values of parameters to get the prior distribution parameters in the same way as it was done in Section~\ref{sect:prior_distr}: $\hat{\mu}_{0,c}=0.749$, ${\sigma_{\mu_0,c}=0.006}$, $ \hat{M}_c=5.064 $, ${\sigma_{M,c}=0.096}$. The posterior distribution is shown in Fig.~\ref{fig:9}. Its expected values are ${\hat{M}_B=5.036}$, ${\hat{\mu}_{0,B}=0.758}$.
	
	The resulting quadrature distribution is presented in Fig.~\ref{fig:4}b together with experimental quadrature histogram and the distribution based on theoretical parameters values. Again, the reconstructed distribution is in a close agreement with expected theoretical one.
	
	Using the Bayesian inference, we measured 30 different states with parameters $ M=1 \div 5 $, $ m=1 $ and $ K=0 \div 5 $. The estimation error rates were from 0.0018\% to 0.287\% for $ \mu_0 $, and from 0.044\% to 0.287\% for $ M $.
	
	\section{Conclusion}
	
	We have considered the problem of statistical parameter estimation of multimode multiphoton subtracted thermal state of light by analyzing the photon statistics as well as the quadrature distribution. We have studied the subsystems containing only a part of the light modes. For statistical estimation we have used a model of photocount distribution (\ref{eq:Pfull}), introduced in \cite{Katamadze2020}. We have shown that the distribution parameters suffer from significant multicollinearity, which complicates its use for unconditional parameter estimation. This task can be simplified if we are able to fix one or several parameters at its true values. However, for a more accurate estimate it is better to variate the prior knowledge. On the basis of the Bayesian approach, it is possible to accurately estimate the photocount distribution parameters. In particular, we were able to reconstruct all the parameters with error rate below 1\% for the sample of size $ n=5.8 \cdot 10^4 $.
	
	Thus, we have developed an approach for the statistical parameters estimation of the multimode states of thermal light with the subtraction of a given number of photons. It can be used to test photon subtraction in multimode states of light. In a similar way, one can solve the problem of characterization of more complex quantum states of light, which have great potential in the quantum computing field. On the other hand, the developed model can also be used to describe single-mode photon subtraction in the case when there is an error in the selection of exactly a single mode.
	
	\section*{Acknowledgments}
	This research was performed according to the Development program of the Interdisciplinary Scientific and Educational School of Lomonosov Moscow State University "Photonic and Quantum technologies. Digital medicine". The work was supported by Russian Science Foundation (RSF), project no: 19-72-10069. GVA expresses his special gratitude to the Russian Foundation for Basic Research (RFBR) for support by the project “PHD students” no: 19-32-90212.

	%\nocite{*}
	\bibliographystyle{apsrev4-2} 
	\bibliography{Article_PRA1}%

\begin{thebibliography}{36}%
\makeatletter
\providecommand \@ifxundefined [1]{%
 \@ifx{#1\undefined}
}%
\providecommand \@ifnum [1]{%
 \ifnum #1\expandafter \@firstoftwo
 \else \expandafter \@secondoftwo
 \fi
}%
\providecommand \@ifx [1]{%
 \ifx #1\expandafter \@firstoftwo
 \else \expandafter \@secondoftwo
 \fi
}%
\providecommand \natexlab [1]{#1}%
\providecommand \enquote  [1]{``#1''}%
\providecommand \bibnamefont  [1]{#1}%
\providecommand \bibfnamefont [1]{#1}%
\providecommand \citenamefont [1]{#1}%
\providecommand \href@noop [0]{\@secondoftwo}%
\providecommand \href [0]{\begingroup \@sanitize@url \@href}%
\providecommand \@href[1]{\@@startlink{#1}\@@href}%
\providecommand \@@href[1]{\endgroup#1\@@endlink}%
\providecommand \@sanitize@url [0]{\catcode `\\12\catcode `\$12\catcode
  `\&12\catcode `\#12\catcode `\^12\catcode `\_12\catcode `\%12\relax}%
\providecommand \@@startlink[1]{}%
\providecommand \@@endlink[0]{}%
\providecommand \url  [0]{\begingroup\@sanitize@url \@url }%
\providecommand \@url [1]{\endgroup\@href {#1}{\urlprefix }}%
\providecommand \urlprefix  [0]{URL }%
\providecommand \Eprint [0]{\href }%
\providecommand \doibase [0]{https://doi.org/}%
\providecommand \selectlanguage [0]{\@gobble}%
\providecommand \bibinfo  [0]{\@secondoftwo}%
\providecommand \bibfield  [0]{\@secondoftwo}%
\providecommand \translation [1]{[#1]}%
\providecommand \BibitemOpen [0]{}%
\providecommand \bibitemStop [0]{}%
\providecommand \bibitemNoStop [0]{.\EOS\space}%
\providecommand \EOS [0]{\spacefactor3000\relax}%
\providecommand \BibitemShut  [1]{\csname bibitem#1\endcsname}%
\let\auto@bib@innerbib\@empty
%</preamble>
\bibitem [{\citenamefont {Ourjoumtsev}(2006)}]{Ourjoumtsev2006}%
  \BibitemOpen
  \bibfield  {author} {\bibinfo {author} {\bibfnamefont {A.}~\bibnamefont
  {Ourjoumtsev}},\ }\href {https://doi.org/10.1126/science.1122858} {\bibfield
  {journal} {\bibinfo  {journal} {Science}\ }\textbf {\bibinfo {volume}
  {312}},\ \bibinfo {pages} {83} (\bibinfo {year} {2006})}\BibitemShut
  {NoStop}%
\bibitem [{\citenamefont {Neergaard-Nielsen}\ \emph {et~al.}(2006)\citenamefont
  {Neergaard-Nielsen}, \citenamefont {Nielsen}, \citenamefont {Hettich},
  \citenamefont {M{\o}lmer},\ and\ \citenamefont
  {Polzik}}]{NeergaardNielsen2006}%
  \BibitemOpen
  \bibfield  {author} {\bibinfo {author} {\bibfnamefont {J.~S.}\ \bibnamefont
  {Neergaard-Nielsen}}, \bibinfo {author} {\bibfnamefont {B.~M.}\ \bibnamefont
  {Nielsen}}, \bibinfo {author} {\bibfnamefont {C.}~\bibnamefont {Hettich}},
  \bibinfo {author} {\bibfnamefont {K.}~\bibnamefont {M{\o}lmer}},\ and\
  \bibinfo {author} {\bibfnamefont {E.~S.}\ \bibnamefont {Polzik}},\ }\href
  {https://doi.org/10.1103/PhysRevLett.97.083604} {\bibfield  {journal}
  {\bibinfo  {journal} {Physical Review Letters}\ }\textbf {\bibinfo {volume}
  {97}},\ \bibinfo {pages} {083604} (\bibinfo {year} {2006})}\BibitemShut
  {NoStop}%
\bibitem [{\citenamefont {Parigi}\ \emph {et~al.}(2007)\citenamefont {Parigi},
  \citenamefont {Zavatta}, \citenamefont {Kim},\ and\ \citenamefont
  {Bellini}}]{Parigi2007}%
  \BibitemOpen
  \bibfield  {author} {\bibinfo {author} {\bibfnamefont {V.}~\bibnamefont
  {Parigi}}, \bibinfo {author} {\bibfnamefont {A.}~\bibnamefont {Zavatta}},
  \bibinfo {author} {\bibfnamefont {M.}~\bibnamefont {Kim}},\ and\ \bibinfo
  {author} {\bibfnamefont {M.}~\bibnamefont {Bellini}},\ }\href
  {https://doi.org/10.1126/science.1146204} {\bibfield  {journal} {\bibinfo
  {journal} {Science}\ }\textbf {\bibinfo {volume} {317}},\ \bibinfo {pages}
  {1890} (\bibinfo {year} {2007})}\BibitemShut {NoStop}%
\bibitem [{\citenamefont {Zavatta}\ \emph {et~al.}(2004)\citenamefont
  {Zavatta}, \citenamefont {Viciani},\ and\ \citenamefont
  {Bellini}}]{Zavatta2004}%
  \BibitemOpen
  \bibfield  {author} {\bibinfo {author} {\bibfnamefont {A.}~\bibnamefont
  {Zavatta}}, \bibinfo {author} {\bibfnamefont {S.}~\bibnamefont {Viciani}},\
  and\ \bibinfo {author} {\bibfnamefont {M.}~\bibnamefont {Bellini}},\ }\href
  {https://doi.org/10.1126/science.1103190} {\bibfield  {journal} {\bibinfo
  {journal} {Science}\ }\textbf {\bibinfo {volume} {306}},\ \bibinfo {pages}
  {660} (\bibinfo {year} {2004})}\BibitemShut {NoStop}%
\bibitem [{\citenamefont {Wenger}\ \emph {et~al.}(2004)\citenamefont {Wenger},
  \citenamefont {Tualle-Brouri},\ and\ \citenamefont {Grangier}}]{Wenger2004}%
  \BibitemOpen
  \bibfield  {author} {\bibinfo {author} {\bibfnamefont {J.}~\bibnamefont
  {Wenger}}, \bibinfo {author} {\bibfnamefont {R.}~\bibnamefont
  {Tualle-Brouri}},\ and\ \bibinfo {author} {\bibfnamefont {P.}~\bibnamefont
  {Grangier}},\ }\href {https://doi.org/10.1103/PhysRevLett.92.153601}
  {\bibfield  {journal} {\bibinfo  {journal} {Physical Review Letters}\
  }\textbf {\bibinfo {volume} {92}},\ \bibinfo {pages} {153601} (\bibinfo
  {year} {2004})}\BibitemShut {NoStop}%
\bibitem [{\citenamefont {Xiang}\ \emph {et~al.}(2010)\citenamefont {Xiang},
  \citenamefont {Ralph}, \citenamefont {Lund}, \citenamefont {Walk},\ and\
  \citenamefont {Pryde}}]{Xiang2010}%
  \BibitemOpen
  \bibfield  {author} {\bibinfo {author} {\bibfnamefont {G.~Y.}\ \bibnamefont
  {Xiang}}, \bibinfo {author} {\bibfnamefont {T.~C.}\ \bibnamefont {Ralph}},
  \bibinfo {author} {\bibfnamefont {A.~P.}\ \bibnamefont {Lund}}, \bibinfo
  {author} {\bibfnamefont {N.}~\bibnamefont {Walk}},\ and\ \bibinfo {author}
  {\bibfnamefont {G.~J.}\ \bibnamefont {Pryde}},\ }\href
  {https://doi.org/10.1038/nphoton.2010.35} {\bibfield  {journal} {\bibinfo
  {journal} {Nature Photonics}\ }\textbf {\bibinfo {volume} {4}},\ \bibinfo
  {pages} {316} (\bibinfo {year} {2010})}\BibitemShut {NoStop}%
\bibitem [{\citenamefont {Costanzo}\ \emph {et~al.}(2017)\citenamefont
  {Costanzo}, \citenamefont {Coelho}, \citenamefont {Biagi}, \citenamefont
  {Fiur{\'{a}}{\v{s}}ek}, \citenamefont {Bellini},\ and\ \citenamefont
  {Zavatta}}]{Costanzo2017}%
  \BibitemOpen
  \bibfield  {author} {\bibinfo {author} {\bibfnamefont {L.~S.}\ \bibnamefont
  {Costanzo}}, \bibinfo {author} {\bibfnamefont {A.~S.}\ \bibnamefont
  {Coelho}}, \bibinfo {author} {\bibfnamefont {N.}~\bibnamefont {Biagi}},
  \bibinfo {author} {\bibfnamefont {J.}~\bibnamefont {Fiur{\'{a}}{\v{s}}ek}},
  \bibinfo {author} {\bibfnamefont {M.}~\bibnamefont {Bellini}},\ and\ \bibinfo
  {author} {\bibfnamefont {A.}~\bibnamefont {Zavatta}},\ }\href
  {https://doi.org/10.1103/PhysRevLett.119.013601} {\bibfield  {journal}
  {\bibinfo  {journal} {Physical Review Letters}\ }\textbf {\bibinfo {volume}
  {119}},\ \bibinfo {pages} {013601} (\bibinfo {year} {2017})}\BibitemShut
  {NoStop}%
\bibitem [{\citenamefont {Katamadze}\ \emph
  {et~al.}(2019{\natexlab{a}})\citenamefont {Katamadze}, \citenamefont
  {Kovlakov}, \citenamefont {Avosopiants},\ and\ \citenamefont
  {Kulik}}]{Katamadze2019a}%
  \BibitemOpen
  \bibfield  {author} {\bibinfo {author} {\bibfnamefont {K.~G.}\ \bibnamefont
  {Katamadze}}, \bibinfo {author} {\bibfnamefont {E.~V.}\ \bibnamefont
  {Kovlakov}}, \bibinfo {author} {\bibfnamefont {G.~V.}\ \bibnamefont
  {Avosopiants}},\ and\ \bibinfo {author} {\bibfnamefont {S.~P.}\ \bibnamefont
  {Kulik}},\ }\href {https://doi.org/10.1364/OL.44.003286} {\bibfield
  {journal} {\bibinfo  {journal} {Optics Letters}\ }\textbf {\bibinfo {volume}
  {44}},\ \bibinfo {pages} {3286} (\bibinfo {year} {2019}{\natexlab{a}})},\
  \Eprint {https://arxiv.org/abs/1901.03093} {arXiv:1901.03093} \BibitemShut
  {NoStop}%
\bibitem [{\citenamefont {Bogdanov}\ \emph {et~al.}(2018)\citenamefont
  {Bogdanov}, \citenamefont {Fastovets}, \citenamefont {Bantysh}, \citenamefont
  {Chernyavskii}, \citenamefont {Semenikhin}, \citenamefont {Bogdanova},
  \citenamefont {Katamadze}, \citenamefont {Kuznetsov}, \citenamefont {Kokin},\
  and\ \citenamefont {Lukichev}}]{Bogdanov2018}%
  \BibitemOpen
  \bibfield  {author} {\bibinfo {author} {\bibfnamefont {{\relax {Yu}}.~I.}\
  \bibnamefont {Bogdanov}}, \bibinfo {author} {\bibfnamefont {D.~V.}\
  \bibnamefont {Fastovets}}, \bibinfo {author} {\bibfnamefont {B.~I.}\
  \bibnamefont {Bantysh}}, \bibinfo {author} {\bibfnamefont {A.~Y.}\
  \bibnamefont {Chernyavskii}}, \bibinfo {author} {\bibfnamefont {I.~A.}\
  \bibnamefont {Semenikhin}}, \bibinfo {author} {\bibfnamefont {N.~A.}\
  \bibnamefont {Bogdanova}}, \bibinfo {author} {\bibfnamefont {K.~G.}\
  \bibnamefont {Katamadze}}, \bibinfo {author} {\bibfnamefont {Y.~A.}\
  \bibnamefont {Kuznetsov}}, \bibinfo {author} {\bibfnamefont {A.~A.}\
  \bibnamefont {Kokin}},\ and\ \bibinfo {author} {\bibfnamefont {V.~F.}\
  \bibnamefont {Lukichev}},\ }\href {https://doi.org/10.1070/QEL16760}
  {\bibfield  {journal} {\bibinfo  {journal} {Quantum Electronics}\ }\textbf
  {\bibinfo {volume} {48}},\ \bibinfo {pages} {1016} (\bibinfo {year}
  {2018})}\BibitemShut {NoStop}%
\bibitem [{\citenamefont {Vidrighin}\ \emph {et~al.}(2016)\citenamefont
  {Vidrighin}, \citenamefont {Dahlsten}, \citenamefont {Barbieri},
  \citenamefont {Kim}, \citenamefont {Vedral},\ and\ \citenamefont
  {Walmsley}}]{Vidrighin2016}%
  \BibitemOpen
  \bibfield  {author} {\bibinfo {author} {\bibfnamefont {M.~D.}\ \bibnamefont
  {Vidrighin}}, \bibinfo {author} {\bibfnamefont {O.}~\bibnamefont {Dahlsten}},
  \bibinfo {author} {\bibfnamefont {M.}~\bibnamefont {Barbieri}}, \bibinfo
  {author} {\bibfnamefont {M.~S.}\ \bibnamefont {Kim}}, \bibinfo {author}
  {\bibfnamefont {V.}~\bibnamefont {Vedral}},\ and\ \bibinfo {author}
  {\bibfnamefont {I.~A.}\ \bibnamefont {Walmsley}},\ }\href
  {https://doi.org/10.1103/PhysRevLett.116.050401} {\bibfield  {journal}
  {\bibinfo  {journal} {Physical Review Letters}\ }\textbf {\bibinfo {volume}
  {116}},\ \bibinfo {pages} {050401} (\bibinfo {year} {2016})}\BibitemShut
  {NoStop}%
\bibitem [{\citenamefont {Hlou{\v{s}}ek}\ \emph {et~al.}(2017)\citenamefont
  {Hlou{\v{s}}ek}, \citenamefont {Je{\v{z}}ek},\ and\ \citenamefont
  {Filip}}]{Hlousek2017}%
  \BibitemOpen
  \bibfield  {author} {\bibinfo {author} {\bibfnamefont {J.}~\bibnamefont
  {Hlou{\v{s}}ek}}, \bibinfo {author} {\bibfnamefont {M.}~\bibnamefont
  {Je{\v{z}}ek}},\ and\ \bibinfo {author} {\bibfnamefont {R.}~\bibnamefont
  {Filip}},\ }\href {https://doi.org/10.1038/s41598-017-13502-0} {\bibfield
  {journal} {\bibinfo  {journal} {Scientific Reports}\ }\textbf {\bibinfo
  {volume} {7}},\ \bibinfo {pages} {13046} (\bibinfo {year}
  {2017})}\BibitemShut {NoStop}%
\bibitem [{\citenamefont {Parazzoli}\ \emph {et~al.}(2016)\citenamefont
  {Parazzoli}, \citenamefont {Capron}, \citenamefont {Koltenbah}, \citenamefont
  {Gerwe}, \citenamefont {Idell}, \citenamefont {Dowling}, \citenamefont
  {Gerry},\ and\ \citenamefont {Boyd}}]{Parazzoli2016}%
  \BibitemOpen
  \bibfield  {author} {\bibinfo {author} {\bibfnamefont {C.~G.}\ \bibnamefont
  {Parazzoli}}, \bibinfo {author} {\bibfnamefont {B.~A.}\ \bibnamefont
  {Capron}}, \bibinfo {author} {\bibfnamefont {B.}~\bibnamefont {Koltenbah}},
  \bibinfo {author} {\bibfnamefont {D.}~\bibnamefont {Gerwe}}, \bibinfo
  {author} {\bibfnamefont {P.}~\bibnamefont {Idell}}, \bibinfo {author}
  {\bibfnamefont {J.}~\bibnamefont {Dowling}}, \bibinfo {author} {\bibfnamefont
  {C.}~\bibnamefont {Gerry}},\ and\ \bibinfo {author} {\bibfnamefont {R.~W.}\
  \bibnamefont {Boyd}},\ }in\ \href
  {https://doi.org/10.1364/CLEO_QELS.2016.FTu3C.4} {\emph {\bibinfo {booktitle}
  {Conference on Lasers and Electro-Optics}}}\ (\bibinfo  {publisher} {OSA},\
  \bibinfo {address} {Washington, D.C.},\ \bibinfo {year} {2016})\ p.\ \bibinfo
  {pages} {FTu3C.4}\BibitemShut {NoStop}%
\bibitem [{\citenamefont {{Hashemi Rafsanjani}}\ \emph
  {et~al.}(2017)\citenamefont {{Hashemi Rafsanjani}}, \citenamefont
  {Mirhosseini}, \citenamefont {Maga{\~{n}}a-Loaiza}, \citenamefont {Gard},
  \citenamefont {Birrittella}, \citenamefont {Koltenbah}, \citenamefont
  {Parazzoli}, \citenamefont {Capron}, \citenamefont {Gerry}, \citenamefont
  {Dowling},\ and\ \citenamefont {Boyd}}]{Boyd2017}%
  \BibitemOpen
  \bibfield  {author} {\bibinfo {author} {\bibfnamefont {S.~M.}\ \bibnamefont
  {{Hashemi Rafsanjani}}}, \bibinfo {author} {\bibfnamefont {M.}~\bibnamefont
  {Mirhosseini}}, \bibinfo {author} {\bibfnamefont {O.~S.}\ \bibnamefont
  {Maga{\~{n}}a-Loaiza}}, \bibinfo {author} {\bibfnamefont {B.~T.}\
  \bibnamefont {Gard}}, \bibinfo {author} {\bibfnamefont {R.}~\bibnamefont
  {Birrittella}}, \bibinfo {author} {\bibfnamefont {B.~E.}\ \bibnamefont
  {Koltenbah}}, \bibinfo {author} {\bibfnamefont {C.~G.}\ \bibnamefont
  {Parazzoli}}, \bibinfo {author} {\bibfnamefont {B.~A.}\ \bibnamefont
  {Capron}}, \bibinfo {author} {\bibfnamefont {C.~C.}\ \bibnamefont {Gerry}},
  \bibinfo {author} {\bibfnamefont {J.~P.}\ \bibnamefont {Dowling}},\ and\
  \bibinfo {author} {\bibfnamefont {R.~W.}\ \bibnamefont {Boyd}},\ }\href
  {https://doi.org/10.1364/OPTICA.4.000487} {\bibfield  {journal} {\bibinfo
  {journal} {Optica}\ }\textbf {\bibinfo {volume} {4}},\ \bibinfo {pages} {487}
  (\bibinfo {year} {2017})}\BibitemShut {NoStop}%
\bibitem [{\citenamefont {Andersen}\ \emph {et~al.}(2015)\citenamefont
  {Andersen}, \citenamefont {Neergaard-Nielsen}, \citenamefont {van Loock},\
  and\ \citenamefont {Furusawa}}]{Andersen2015}%
  \BibitemOpen
  \bibfield  {author} {\bibinfo {author} {\bibfnamefont {U.~L.}\ \bibnamefont
  {Andersen}}, \bibinfo {author} {\bibfnamefont {J.~S.}\ \bibnamefont
  {Neergaard-Nielsen}}, \bibinfo {author} {\bibfnamefont {P.}~\bibnamefont {van
  Loock}},\ and\ \bibinfo {author} {\bibfnamefont {A.}~\bibnamefont
  {Furusawa}},\ }\href {https://doi.org/10.1038/nphys3410} {\bibfield
  {journal} {\bibinfo  {journal} {Nature Physics}\ }\textbf {\bibinfo {volume}
  {11}},\ \bibinfo {pages} {713} (\bibinfo {year} {2015})}\BibitemShut
  {NoStop}%
\bibitem [{\citenamefont {Ra}\ \emph {et~al.}(2020)\citenamefont {Ra},
  \citenamefont {Dufour}, \citenamefont {Walschaers}, \citenamefont {Jacquard},
  \citenamefont {Michel}, \citenamefont {Fabre},\ and\ \citenamefont
  {Treps}}]{Ra2020}%
  \BibitemOpen
  \bibfield  {author} {\bibinfo {author} {\bibfnamefont {Y.-S.}\ \bibnamefont
  {Ra}}, \bibinfo {author} {\bibfnamefont {A.}~\bibnamefont {Dufour}}, \bibinfo
  {author} {\bibfnamefont {M.}~\bibnamefont {Walschaers}}, \bibinfo {author}
  {\bibfnamefont {C.}~\bibnamefont {Jacquard}}, \bibinfo {author}
  {\bibfnamefont {T.}~\bibnamefont {Michel}}, \bibinfo {author} {\bibfnamefont
  {C.}~\bibnamefont {Fabre}},\ and\ \bibinfo {author} {\bibfnamefont
  {N.}~\bibnamefont {Treps}},\ }\href
  {https://doi.org/10.1038/s41567-019-0726-y} {\bibfield  {journal} {\bibinfo
  {journal} {Nature Physics}\ }\textbf {\bibinfo {volume} {16}},\ \bibinfo
  {pages} {144} (\bibinfo {year} {2020})}\BibitemShut {NoStop}%
\bibitem [{\citenamefont {Ra}\ \emph {et~al.}(2017)\citenamefont {Ra},
  \citenamefont {Jacquard}, \citenamefont {Dufour}, \citenamefont {Fabre},\
  and\ \citenamefont {Treps}}]{Ra2017}%
  \BibitemOpen
  \bibfield  {author} {\bibinfo {author} {\bibfnamefont {Y.-S.}\ \bibnamefont
  {Ra}}, \bibinfo {author} {\bibfnamefont {C.}~\bibnamefont {Jacquard}},
  \bibinfo {author} {\bibfnamefont {A.}~\bibnamefont {Dufour}}, \bibinfo
  {author} {\bibfnamefont {C.}~\bibnamefont {Fabre}},\ and\ \bibinfo {author}
  {\bibfnamefont {N.}~\bibnamefont {Treps}},\ }\href
  {https://doi.org/10.1103/PhysRevX.7.031012} {\bibfield  {journal} {\bibinfo
  {journal} {Physical Review X}\ }\textbf {\bibinfo {volume} {7}},\ \bibinfo
  {pages} {031012} (\bibinfo {year} {2017})}\BibitemShut {NoStop}%
\bibitem [{\citenamefont {Agarwal}(1992)}]{Agarwal1992}%
  \BibitemOpen
  \bibfield  {author} {\bibinfo {author} {\bibfnamefont {G.~S.}\ \bibnamefont
  {Agarwal}},\ }\href {https://doi.org/10.1103/PhysRevA.45.1787} {\bibfield
  {journal} {\bibinfo  {journal} {Physical Review A}\ }\textbf {\bibinfo
  {volume} {45}},\ \bibinfo {pages} {1787} (\bibinfo {year}
  {1992})}\BibitemShut {NoStop}%
\bibitem [{\citenamefont {Allevi}\ \emph {et~al.}(2010)\citenamefont {Allevi},
  \citenamefont {Andreoni}, \citenamefont {Bondani}, \citenamefont {Genoni},\
  and\ \citenamefont {Olivares}}]{Allevi2010}%
  \BibitemOpen
  \bibfield  {author} {\bibinfo {author} {\bibfnamefont {A.}~\bibnamefont
  {Allevi}}, \bibinfo {author} {\bibfnamefont {A.}~\bibnamefont {Andreoni}},
  \bibinfo {author} {\bibfnamefont {M.}~\bibnamefont {Bondani}}, \bibinfo
  {author} {\bibfnamefont {M.~G.}\ \bibnamefont {Genoni}},\ and\ \bibinfo
  {author} {\bibfnamefont {S.}~\bibnamefont {Olivares}},\ }\href
  {https://doi.org/10.1103/PhysRevA.82.013816} {\bibfield  {journal} {\bibinfo
  {journal} {Physical Review A}\ }\textbf {\bibinfo {volume} {82}},\ \bibinfo
  {pages} {013816} (\bibinfo {year} {2010})}\BibitemShut {NoStop}%
\bibitem [{\citenamefont {Zhai}\ \emph {et~al.}(2013)\citenamefont {Zhai},
  \citenamefont {Becerra}, \citenamefont {Glebov}, \citenamefont {Wen},
  \citenamefont {Lita}, \citenamefont {Calkins}, \citenamefont {Gerrits},
  \citenamefont {Fan}, \citenamefont {Nam},\ and\ \citenamefont
  {Migdall}}]{Zhai2013}%
  \BibitemOpen
  \bibfield  {author} {\bibinfo {author} {\bibfnamefont {Y.}~\bibnamefont
  {Zhai}}, \bibinfo {author} {\bibfnamefont {F.~E.}\ \bibnamefont {Becerra}},
  \bibinfo {author} {\bibfnamefont {B.~L.}\ \bibnamefont {Glebov}}, \bibinfo
  {author} {\bibfnamefont {J.}~\bibnamefont {Wen}}, \bibinfo {author}
  {\bibfnamefont {A.~E.}\ \bibnamefont {Lita}}, \bibinfo {author}
  {\bibfnamefont {B.}~\bibnamefont {Calkins}}, \bibinfo {author} {\bibfnamefont
  {T.}~\bibnamefont {Gerrits}}, \bibinfo {author} {\bibfnamefont
  {J.}~\bibnamefont {Fan}}, \bibinfo {author} {\bibfnamefont {S.~W.}\
  \bibnamefont {Nam}},\ and\ \bibinfo {author} {\bibfnamefont {A.}~\bibnamefont
  {Migdall}},\ }\href {https://doi.org/10.1364/OL.38.002171} {\bibfield
  {journal} {\bibinfo  {journal} {Optics Letters}\ }\textbf {\bibinfo {volume}
  {38}},\ \bibinfo {pages} {2171} (\bibinfo {year} {2013})}\BibitemShut
  {NoStop}%
\bibitem [{\citenamefont {Bogdanov}\ \emph {et~al.}(2017)\citenamefont
  {Bogdanov}, \citenamefont {Katamadze}, \citenamefont {Avosopiants},
  \citenamefont {Belinsky}, \citenamefont {Bogdanova}, \citenamefont
  {Kalinkin},\ and\ \citenamefont {Kulik}}]{Bogdanov2017}%
  \BibitemOpen
  \bibfield  {author} {\bibinfo {author} {\bibfnamefont {{\relax Yu}.~I.}\
  \bibnamefont {Bogdanov}}, \bibinfo {author} {\bibfnamefont {K.~G.}\
  \bibnamefont {Katamadze}}, \bibinfo {author} {\bibfnamefont {G.~V.}\
  \bibnamefont {Avosopiants}}, \bibinfo {author} {\bibfnamefont {L.~V.}\
  \bibnamefont {Belinsky}}, \bibinfo {author} {\bibfnamefont {N.~A.}\
  \bibnamefont {Bogdanova}}, \bibinfo {author} {\bibfnamefont {A.~A.}\
  \bibnamefont {Kalinkin}},\ and\ \bibinfo {author} {\bibfnamefont {S.~P.}\
  \bibnamefont {Kulik}},\ }\href {https://doi.org/10.1103/PhysRevA.96.063803}
  {\bibfield  {journal} {\bibinfo  {journal} {Physical Review A}\ }\textbf
  {\bibinfo {volume} {96}},\ \bibinfo {pages} {063803} (\bibinfo {year}
  {2017})}\BibitemShut {NoStop}%
\bibitem [{\citenamefont {Bogdanov}\ \emph {et~al.}(2003)\citenamefont
  {Bogdanov}, \citenamefont {Bogdanova},\ and\ \citenamefont
  {Dshkhunyan}}]{Bogdanov2003}%
  \BibitemOpen
  \bibfield  {author} {\bibinfo {author} {\bibfnamefont {{\relax {Yu}}.~I.}\
  \bibnamefont {Bogdanov}}, \bibinfo {author} {\bibfnamefont {N.~A.}\
  \bibnamefont {Bogdanova}},\ and\ \bibinfo {author} {\bibfnamefont {V.~L.}\
  \bibnamefont {Dshkhunyan}},\ }\href@noop {} {\bibfield  {journal} {\bibinfo
  {journal} {Russian Microelectronics}\ }\textbf {\bibinfo {volume} {32}},\
  \bibinfo {pages} {51} (\bibinfo {year} {2003})}\BibitemShut {NoStop}%
\bibitem [{\citenamefont {Bogdanov}\ \emph
  {et~al.}(2016{\natexlab{a}})\citenamefont {Bogdanov}, \citenamefont
  {Bogdanova}, \citenamefont {Katamadze}, \citenamefont {Avosopyants},\ and\
  \citenamefont {Lukichev}}]{Bogdanov2016a}%
  \BibitemOpen
  \bibfield  {author} {\bibinfo {author} {\bibfnamefont {{\relax {Yu}}.~I.}\
  \bibnamefont {Bogdanov}}, \bibinfo {author} {\bibfnamefont {N.~A.}\
  \bibnamefont {Bogdanova}}, \bibinfo {author} {\bibfnamefont {K.~G.}\
  \bibnamefont {Katamadze}}, \bibinfo {author} {\bibfnamefont {G.~V.}\
  \bibnamefont {Avosopyants}},\ and\ \bibinfo {author} {\bibfnamefont {V.~F.}\
  \bibnamefont {Lukichev}},\ }\href {https://doi.org/10.3103/S8756699016050095}
  {\bibfield  {journal} {\bibinfo  {journal} {Optoelectronics, Instrumentation
  and Data Processing}\ }\textbf {\bibinfo {volume} {52}},\ \bibinfo {pages}
  {475} (\bibinfo {year} {2016}{\natexlab{a}})}\BibitemShut {NoStop}%
\bibitem [{\citenamefont {Katamadze}\ \emph
  {et~al.}(2019{\natexlab{b}})\citenamefont {Katamadze}, \citenamefont
  {Avosopiants}, \citenamefont {Bantysh}, \citenamefont {Bogdanov},\ and\
  \citenamefont {Kulik}}]{Katamadze2019}%
  \BibitemOpen
  \bibfield  {author} {\bibinfo {author} {\bibfnamefont {K.~G.}\ \bibnamefont
  {Katamadze}}, \bibinfo {author} {\bibfnamefont {G.~V.}\ \bibnamefont
  {Avosopiants}}, \bibinfo {author} {\bibfnamefont {B.~I.}\ \bibnamefont
  {Bantysh}}, \bibinfo {author} {\bibfnamefont {{\relax {Yu}}.~I.}\
  \bibnamefont {Bogdanov}},\ and\ \bibinfo {author} {\bibfnamefont {S.~P.}\
  \bibnamefont {Kulik}},\ }in\ \href {https://doi.org/10.1117/12.2522080}
  {\emph {\bibinfo {booktitle} {International Conference on Micro- and
  Nano-Electronics 2018}}},\ Vol.\ \bibinfo {volume} {11022},\ \bibinfo
  {editor} {edited by\ \bibinfo {editor} {\bibfnamefont {V.~F.}\ \bibnamefont
  {Lukichev}}\ and\ \bibinfo {editor} {\bibfnamefont {K.~V.}\ \bibnamefont
  {Rudenko}}}\ (\bibinfo  {publisher} {SPIE},\ \bibinfo {year} {2019})\ p.\
  \bibinfo {pages} {110222K}\BibitemShut {NoStop}%
\bibitem [{\citenamefont {{L. Mandel and E. Wolf}}(1995)}]{Mandel1995}%
  \BibitemOpen
  \bibfield  {author} {\bibinfo {author} {\bibnamefont {{L. Mandel and E.
  Wolf}}},\ }\href@noop {} {\emph {\bibinfo {title} {{Optical Coherence and
  Quantum Optics}}}},\ \bibinfo {edition} {1st}\ ed.\ (\bibinfo  {publisher}
  {Cambridge University Press},\ \bibinfo {year} {1995})\ p.\ \bibinfo {pages}
  {1194}\BibitemShut {NoStop}%
\bibitem [{\citenamefont {Bogdanov}\ \emph {et~al.}(2019)\citenamefont
  {Bogdanov}, \citenamefont {Bogdanova}, \citenamefont {Katamadze},
  \citenamefont {Avosopiants},\ and\ \citenamefont {Kulik}}]{Avosopiants2019}%
  \BibitemOpen
  \bibfield  {author} {\bibinfo {author} {\bibfnamefont {{\relax {Yu}}.~I.}\
  \bibnamefont {Bogdanov}}, \bibinfo {author} {\bibfnamefont {N.~A.}\
  \bibnamefont {Bogdanova}}, \bibinfo {author} {\bibfnamefont {K.~G.}\
  \bibnamefont {Katamadze}}, \bibinfo {author} {\bibfnamefont {G.~V.}\
  \bibnamefont {Avosopiants}},\ and\ \bibinfo {author} {\bibfnamefont {S.~P.}\
  \bibnamefont {Kulik}},\ }in\ \href {https://doi.org/10.1117/12.2521910}
  {\emph {\bibinfo {booktitle} {International Conference on Micro- and
  Nano-Electronics 2018}}},\ \bibinfo {editor} {edited by\ \bibinfo {editor}
  {\bibfnamefont {V.~F.}\ \bibnamefont {Lukichev}}\ and\ \bibinfo {editor}
  {\bibfnamefont {K.~V.}\ \bibnamefont {Rudenko}}}\ (\bibinfo  {publisher}
  {SPIE},\ \bibinfo {year} {2019})\ p.\ \bibinfo {pages} {110222L}\BibitemShut
  {NoStop}%
\bibitem [{\citenamefont {Katamadze}\ \emph {et~al.}(2020)\citenamefont
  {Katamadze}, \citenamefont {Avosopiants}, \citenamefont {Bogdanova},
  \citenamefont {Bogdanov},\ and\ \citenamefont {Kulik}}]{Katamadze2020}%
  \BibitemOpen
  \bibfield  {author} {\bibinfo {author} {\bibfnamefont {K.~G.}\ \bibnamefont
  {Katamadze}}, \bibinfo {author} {\bibfnamefont {G.~V.}\ \bibnamefont
  {Avosopiants}}, \bibinfo {author} {\bibfnamefont {N.~A.}\ \bibnamefont
  {Bogdanova}}, \bibinfo {author} {\bibfnamefont {{\relax {Yu}}.~I.}\
  \bibnamefont {Bogdanov}},\ and\ \bibinfo {author} {\bibfnamefont {S.~P.}\
  \bibnamefont {Kulik}},\ }\href {https://doi.org/10.1103/PhysRevA.101.013811}
  {\bibfield  {journal} {\bibinfo  {journal} {Physical Review A}\ }\textbf
  {\bibinfo {volume} {101}},\ \bibinfo {pages} {013811} (\bibinfo {year}
  {2020})}\BibitemShut {NoStop}%
\bibitem [{\citenamefont {Landau}\ and\ \citenamefont
  {Lifshitz}(1959)}]{Landau1959}%
  \BibitemOpen
  \bibfield  {author} {\bibinfo {author} {\bibfnamefont {L.}~\bibnamefont
  {Landau}}\ and\ \bibinfo {author} {\bibfnamefont {E.}~\bibnamefont
  {Lifshitz}},\ }\href@noop {} {\emph {\bibinfo {title} {{Statistical Physics.
  Part 1}}}},\ \bibinfo {edition} {1st}\ ed.\ (\bibinfo  {publisher} {Pergamon
  Press},\ \bibinfo {year} {1959})\ p.\ \bibinfo {pages} {484}\BibitemShut
  {NoStop}%
\bibitem [{\citenamefont {Feller}(1968)}]{Feller1968}%
  \BibitemOpen
  \bibfield  {author} {\bibinfo {author} {\bibfnamefont {W.}~\bibnamefont
  {Feller}},\ }\href@noop {} {\emph {\bibinfo {title} {{An Introduction to
  Probability Theory and Its Applications}}}},\ \bibinfo {edition} {3rd}\ ed.\
  (\bibinfo  {publisher} {Wiley},\ \bibinfo {year} {1968})\ p.\ \bibinfo
  {pages} {524}\BibitemShut {NoStop}%
\bibitem [{\citenamefont {Martienssen}(1964)}]{Martienssen1964}%
  \BibitemOpen
  \bibfield  {author} {\bibinfo {author} {\bibfnamefont {W.}~\bibnamefont
  {Martienssen}},\ }\href {https://doi.org/10.1119/1.1970023} {\bibfield
  {journal} {\bibinfo  {journal} {American Journal of Physics}\ }\textbf
  {\bibinfo {volume} {32}},\ \bibinfo {pages} {919} (\bibinfo {year}
  {1964})}\BibitemShut {NoStop}%
\bibitem [{\citenamefont {Arecchi}(1965)}]{Arecchi1965}%
  \BibitemOpen
  \bibfield  {author} {\bibinfo {author} {\bibfnamefont {F.~T.}\ \bibnamefont
  {Arecchi}},\ }\href {https://doi.org/10.1103/PhysRevLett.15.912} {\bibfield
  {journal} {\bibinfo  {journal} {Physical Review Letters}\ }\textbf {\bibinfo
  {volume} {15}},\ \bibinfo {pages} {912} (\bibinfo {year} {1965})}\BibitemShut
  {NoStop}%
\bibitem [{\citenamefont {Fisher}(1935)}]{Fisher1935}%
  \BibitemOpen
  \bibfield  {author} {\bibinfo {author} {\bibfnamefont {R.~A.}\ \bibnamefont
  {Fisher}},\ }\href {https://doi.org/10.1111/j.1469-1809.1935.tb02120.x}
  {\bibfield  {journal} {\bibinfo  {journal} {Annals of Eugenics}\ }\textbf
  {\bibinfo {volume} {6}},\ \bibinfo {pages} {391} (\bibinfo {year}
  {1935})}\BibitemShut {NoStop}%
\bibitem [{\citenamefont {Cox}(2006)}]{Cox2006}%
  \BibitemOpen
  \bibfield  {author} {\bibinfo {author} {\bibfnamefont {D.~R.}\ \bibnamefont
  {Cox}},\ }\href {https://doi.org/10.1017/CBO9780511813559} {\emph {\bibinfo
  {title} {{Principles of Statistical Inference}}}},\ \bibinfo {edition} {1st}\
  ed.\ (\bibinfo  {publisher} {Cambridge University Press},\ \bibinfo {address}
  {Cambridge},\ \bibinfo {year} {2006})\ p.\ \bibinfo {pages} {236}\BibitemShut
  {NoStop}%
\bibitem [{\citenamefont {Kendall}\ and\ \citenamefont
  {Stuart}(1961)}]{Kendall1961}%
  \BibitemOpen
  \bibfield  {author} {\bibinfo {author} {\bibfnamefont {M.~G.}\ \bibnamefont
  {Kendall}}\ and\ \bibinfo {author} {\bibfnamefont {A.}~\bibnamefont
  {Stuart}},\ }\href@noop {} {\emph {\bibinfo {title} {{The Advanced Theory of
  Statistics, Vol. 2: Inference and Relationship}}}}\ (\bibinfo  {publisher}
  {Charles Griffin {\&} Company Ltd.},\ \bibinfo {address} {London},\ \bibinfo
  {year} {1961})\BibitemShut {NoStop}%
\bibitem [{\citenamefont {Gelman}\ \emph {et~al.}(2013)\citenamefont {Gelman},
  \citenamefont {Carlin}, \citenamefont {Stern}, \citenamefont {Dunson},
  \citenamefont {Vehtari},\ and\ \citenamefont {Rubin}}]{Gelman2013}%
  \BibitemOpen
  \bibfield  {author} {\bibinfo {author} {\bibfnamefont {A.}~\bibnamefont
  {Gelman}}, \bibinfo {author} {\bibfnamefont {J.}~\bibnamefont {Carlin}},
  \bibinfo {author} {\bibfnamefont {H.}~\bibnamefont {Stern}}, \bibinfo
  {author} {\bibfnamefont {D.}~\bibnamefont {Dunson}}, \bibinfo {author}
  {\bibfnamefont {A.}~\bibnamefont {Vehtari}},\ and\ \bibinfo {author}
  {\bibfnamefont {D.}~\bibnamefont {Rubin}},\ }\href@noop {} {\emph {\bibinfo
  {title} {{Bayesian Data Analysis}}}},\ \bibinfo {edition} {3rd}\ ed.\
  (\bibinfo  {publisher} {Chapman and Hall/CRC Press},\ \bibinfo {year}
  {2013})\ p.\ \bibinfo {pages} {675}\BibitemShut {NoStop}%
\bibitem [{\citenamefont {Leonhardt}(1997)}]{Leonhardt1997}%
  \BibitemOpen
  \bibfield  {author} {\bibinfo {author} {\bibfnamefont {U.}~\bibnamefont
  {Leonhardt}},\ }\href@noop {} {\emph {\bibinfo {title} {{Measuring the
  Quantum State of Light}}}},\ \bibinfo {edition} {1st}\ ed.\ (\bibinfo
  {publisher} {Cambridge University Press},\ \bibinfo {year} {1997})\ p.\
  \bibinfo {pages} {208}\BibitemShut {NoStop}%
\bibitem [{\citenamefont {Bogdanov}\ \emph
  {et~al.}(2016{\natexlab{b}})\citenamefont {Bogdanov}, \citenamefont
  {Avosopyants}, \citenamefont {Belinskii}, \citenamefont {Katamadze},
  \citenamefont {Kulik},\ and\ \citenamefont {Lukichev}}]{Bogdanov2016}%
  \BibitemOpen
  \bibfield  {author} {\bibinfo {author} {\bibfnamefont {{\relax {Yu}}.~I.}\
  \bibnamefont {Bogdanov}}, \bibinfo {author} {\bibfnamefont {G.~V.}\
  \bibnamefont {Avosopyants}}, \bibinfo {author} {\bibfnamefont {L.~V.}\
  \bibnamefont {Belinskii}}, \bibinfo {author} {\bibfnamefont {K.~G.}\
  \bibnamefont {Katamadze}}, \bibinfo {author} {\bibfnamefont {S.~P.}\
  \bibnamefont {Kulik}},\ and\ \bibinfo {author} {\bibfnamefont {V.~F.}\
  \bibnamefont {Lukichev}},\ }\href {https://doi.org/10.1134/S1063776116070025}
  {\bibfield  {journal} {\bibinfo  {journal} {Journal of Experimental and
  Theoretical Physics}\ }\textbf {\bibinfo {volume} {123}},\ \bibinfo {pages}
  {212} (\bibinfo {year} {2016}{\natexlab{b}})}\BibitemShut {NoStop}%
\end{thebibliography}%
	
\end{document}